\documentclass[review]{elsarticle}
\usepackage{graphicx}
\usepackage{amsmath}
\usepackage{amssymb}
\usepackage{bm}

\usepackage{ifthen}
\usepackage{multimedia}

\usepackage{animate}

\newcounter{isElectronicVersion} 

\DeclareMathOperator\dv{div} 
\DeclareMathOperator\id{I}   
\DeclareMathOperator\tr{tr}   
\DeclareMathOperator\sgn{sgn}   
\newcommand{\bu}{\mathbf{u}}  
\newcommand{\bq}{\mathbf{q}}  
\newcommand{\bx}{\mathbf{x}}  
\newcommand{\be}{\mathbf{e}}  
\newcommand{\bbs}{\mathbf{s}}  
\newcommand{\bn}{\mathbf{n}}  
\newcommand{\mx}{\mathrm{max}}  
\newcommand{\mn}{\mathrm{min}}  

\newcommand{\bsigma}{\bm{\sigma}} 
\newcommand{\btau}{\bm{\tau}} 
\newcommand{\epss}{{\cal E}}

\newcommand{\pd}[2]{ 
	\dfrac{\partial #1}{\partial #2}
}

\newtheorem{dfn}{Definition}

\graphicspath{{img/}}

\begin{document}	
	\title{Non-symmetry of a hydraulic fracture due to the inhomogeneity of the reservoir}
	
	\author{A.N. Baykin$^{1,2}$, S.V. Golovin$^{1,2}$}
	
	\address{$^1$Lavrentyev Institute of Hydrodynamics, Novosibirsk, Russia\\
		$^2$Novosibirsk State University, Novosibirsk, Russia}
	
	\ead{alexey.baykin@gmail.com, golovin@hydro.nsc.ru}
	
	\begin{abstract}
It is usually assumed that hydraulic fracture has two symmetrical wings with respect to the fluid injection point. Hence, authors limit themselves to modelling of one half of the fracture. In our work we demonstrate that the case of a symmetrical fracture occurs only in a homogeneous reservoir with constant physical parameters and confining {\it in situ} stress. Otherwise, inhomogeneity in the stress or the rock permeability can significantly change the dynamics of the fracture propagation. The mathematical model of the hydraulic fracturing used in the paper is adopted from our earlier work  \cite{Golovin_Baykin_2016_Pore}. In present paper we perform numerical experiments demonstrating that in case of non-constant confining stress or reservoir permeability the fracture is developing non-symmetrically. An important role is played by the action of the backstress that is formed near the fracture due to the pore pressure. To support this observation we give a formal definition of the backstress and compare the values of the backstress acting on the two different fracture wings. These numerical experiments underline the importance of the proper modelling of the interaction between pore and fracturing fluids for the correct simulation of the hydraulic fracturing.
	\end{abstract}
\maketitle
\section{Introduction}
Hydraulic fracturing is the process of creating a fracture due to pumping a highly pressurized fluid in a rock formation through a wellbore that is commonly used for intensification of hydrocarbon production. Widespread use of hydraulic fracturing in low-permeable reservoirs explains the demand for mathematical modelling of the fracturing process. Compared to the classical models of brittle fractures, hydraulic fracture theory describes the fluid flow within the fracture coupled with the fracture opening and extention, exchange of the fracturing fluid with the pore fluid, influence of the pore pressure on the stresses, etc. 

The recent progress in modelling of hydraulic fracture dynamics has been described in review papers \cite{Adachi, Detournay2016} and citations therein. The complexity of the hydraulic fracturing process and the requirement that the model should be ready for engineering use (i.e., it should give a result within a reasonable calculation time on a personal computer) leads to the necessity for various simplification assumptions. For example, the widely used models of Khristianovich, Zheltov, Geertsma and de Klerk (KGD) \cite{Z-K, GeertsmadeKlerk1969} and Perkins, Kern and Nordgen (PKN) \cite{PerkinsKern1961, Nordgren1972} assume that the fracture is propagating in infinite elastic medium, that no influence of pore pressure is taken into account and that the fracture propagates along a straight path such that it has two symmetrical wings relative to the point of fluid injection. The same symmetry assumption is used in most modern papers that are based on Biot's poroelasticity model \cite{Carrier_Granet}. 

Although the assumption of the fracture symmetry is natural in many situations, it can be violated by a non-constant confining stress or inhomogeneous physical parameters of the reservoir. The goal of the present work is to perform a sensitivity analysis of the fracture dynamics to the above-mentioned factors. In this paper, we use a numerical model that describes the propagation of a hydraulic fracture in an inhomogeneous poroelastic medium subjected to the non-constant confining stresses proposed in  \cite{Golovin_Baykin_2016_Pore}. The model is based on Biot's poroelasticity equations  \cite{Biot1955, Biot1965}, supplemented by the lubrication equation for fluid flow within the fracture. An exchange of fluid between the fracture and the reservoir is described by the boundary conditions over the fracture's walls. The correct interactions between the filtration, pore pressure and stresses within the reservoir are guaranteed by the corresponding terms in the stress tensor and in the model equations. 

At first, we show that the non-uniformness of the confining {\it in situ} stress leads to a non-symmetrical fracture. The fracture propagates in the direction of the lower {\it in situ} stress even for a very small stress contrasts that are of the order of 1\%. Next, we demonstrate that the non-uniform confining stress acting on the fracture's walls can be provoked by the difference of the backstress due to the inhomogeneity of the physical properties of the reservoir. Indeed, the difference in the permeability of the rock leads to the variability of the fluid filtrations into the reservoir. In turn, this generates a difference in the backstress due to the action of the pore pressure. In order to quantify its influence, we give a formal definition of the backstress and calculate its value over both wings of the fracture. It is shown that the backstress can be of the same order as the {\it in situ} stress and, thus, can modify the fracture dynamics. Basing on the series of the numerical experiments we make the conclusion that the account for the inhomogeneity of physical properties of the reservoir as well as the proper modelling of the interaction between the pore pressure and elastic stress within the reservoir are important for the correct modelling of the hydraulic fracturing.

\section{Mathematical formulation of the problem} \label{sec:problem_formul}

The poroelastic rock is described in terms of the porosity $\phi$ and the permeability $k_r(\bx)$, with the solid phase displacement $\bu(t,\bx)$. Pores are saturated by a single-phase Newtonian fluid with the effective viscosity $\eta_r$. The pore pressure is denoted as $p(t,\bx)$. We use the linear Darcy law for the fluid velocity $\bq=-(k_r/\eta_r)\nabla p$. It is supposed that fracturing fluid is also Newtonian with different viscosity $\eta_f$. However, it is assumed that at the process of filtration of fracturing fluid to the reservoir, the infiltrated fluid has the same viscosity as the original pore fluid. This assumption is motivated by the observation that usually the fracturing fluid is a high-viscous gel and only its low-viscous base fluid is filtrated into the reservoir. 

The vertical planar hydraulic fracture is supposed to have a constant width along the vertical coordinate so that the plain strain approximation is applicable. We choose the coordinate system such that $x$-axis coincides with the horizontal direction of fracture propagation and $z$-axis coincides with the vertical direction. Fluid is injected into the fracture through the vertical wellbore that is located at $x=0$ such that $z$-axis is located in the centre of the wellbore. We suppose that the fracture's aperture is equal to $2w(t,x)$. According to our assumptions the plain strain approximation is applicable so that we can limit ourselves to the 2D problem in the cross-section by plane $z=0$.

\begin{figure}[t]
		\centering
		\includegraphics[width=0.6\linewidth]{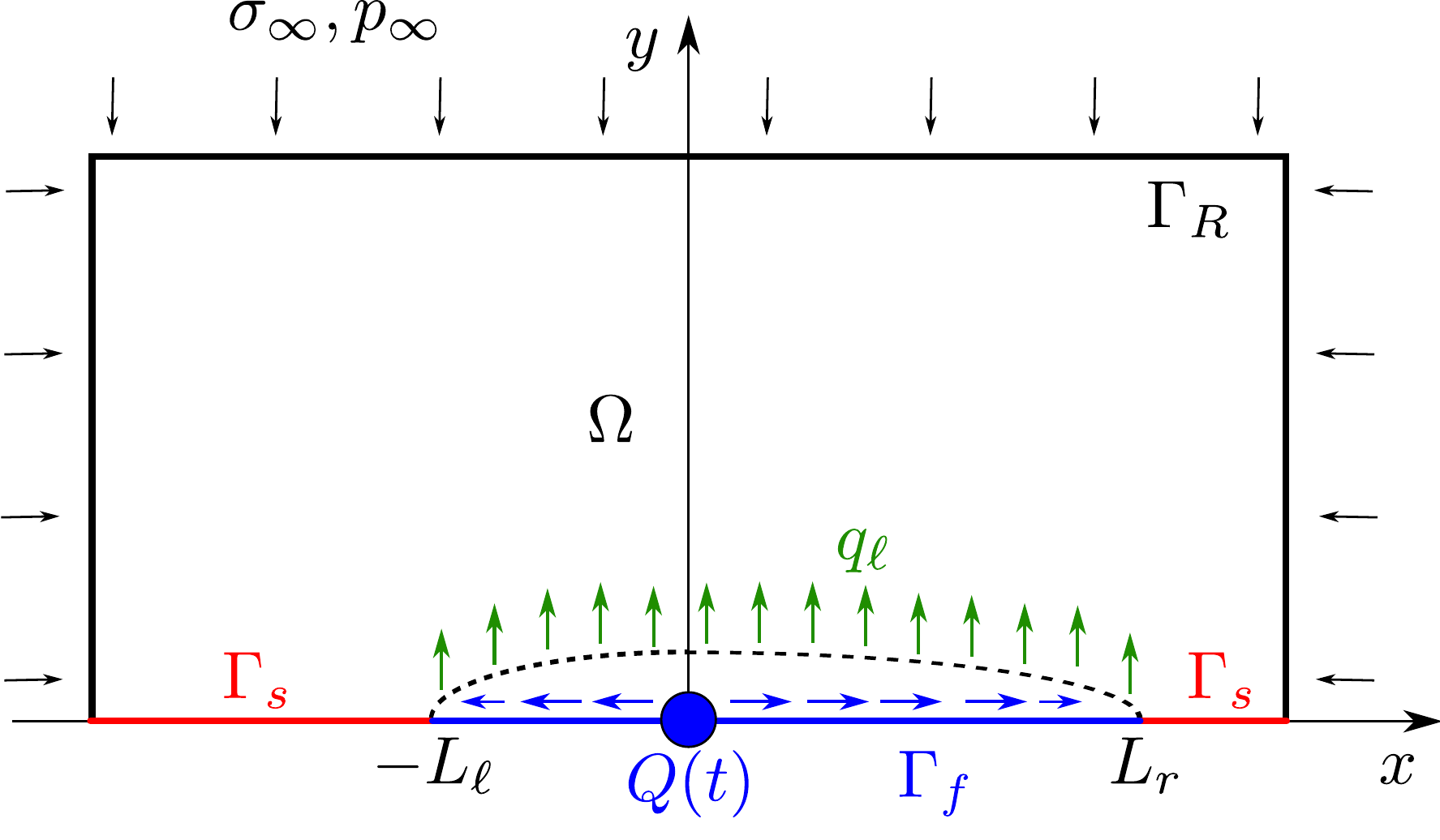}
		\caption{The schematic view of the modelling area comprising the horizontal cross-section of the fracture by plane $z=0$}
		\label{fig:fracture_geometry_2d}
\end{figure}
	
The governing equation of the quasi-static poroelasticity model can be written as:
	\begin{equation}\label{eq:model}
        \begin{array}{l}
            \displaystyle \dv{\btau}=0, \quad \btau = \btau^0 + \lambda \dv{\bu} \id+2\mu\, \epss{(\bu)}-\alpha p \id,\\[4mm]
            \displaystyle S_\varepsilon\pd{p}{t} = \dv{\Bigl( \frac{k_r}{\eta_r}\nabla p - \alpha \pd{\bu}{t} \Bigr)}.
        \end{array}
    \end{equation}
Here $\epss(\bu)$ is the Cauchy's strain tensor
$2\epss(\bu)_{ij}=\partial u_i/\partial x_j+\partial u_j/\partial x_i$ $(i,j=1,2)$, $\alpha$ is the Biot's coefficient, $\lambda(\bx)$ and $\mu(\bx)$ are  elasticity moduli, $\id$ is the identity tensor, $\btau_0(\bx)$ is a tensor of the initial prestress due to the {\it in situ} formation geological stress. The storativity $S_\varepsilon$ reflects the dependence of the Lagrangian porosity  $\phi$ on $\epsilon = \tr\epss$ and $p$ as \cite{Coussy}:
	\begin{equation}\label{eq:storativity_def}    
	\pd{\phi}{t}=\alpha \pd{\epsilon}{t}+S_\varepsilon\pd{p}{t},\quad 		S_\varepsilon = \dfrac{(\phi_0-\alpha)(1-\alpha)}{K},
	\end{equation} 
where $K = \lambda + \dfrac{2\mu}{3}$ is the bulk modulus, $\phi_0$ is the initial porosity. Due to the plane strain approximation, the solid phase displacement vector $\bu = (u_1, u_2) = (u, v)$
is two-dimensional, all vector operations are also taken in 2D space of independent variables $\bx=(x_1, x_2) = (x, y)$.
	
Symmetry of the problem with respect to $Ox$-axis allows solving equations \eqref{eq:model} in domain $\Omega = \{(x,y): |x|\leq R, 0\leq y\leq R\}$ as shown in Figure \ref{fig:fracture_geometry_2d}.

Over the outer boundary $\Gamma_R=\{\partial \Omega\cap y>0\}$ the 
confining far-field stress  $\bsigma_\infty$ is applied and the  constant pore pressure $p=p_\infty$ is prescribed:
\begin{equation}\label{eq:GammaRBC}
	    \Gamma_R:\quad p=p_\infty,\quad \btau\langle\bn\rangle=\bsigma_\infty,
	    \quad (\btau\langle\bn\rangle)_i=\tau_{ij}n_j.
\end{equation}

Henceforth $\bn$ and $\bbs$ denote the outer normal and tangential unit  vectors to the boundary of the domain $\Omega$; the summation over the repeating indices is implied.  In this paper we regard the case
\[\bsigma_\infty = 
\left\{
\begin{array}{ll}
-\sigma_\infty^\mn \be_2, & |x|\le R,\quad y=R\\[2mm]
-\sigma_\infty^\mx \sgn(x)\be_1, & |x|=R, \quad 0\le y\le R.
\end{array}\right.\] 
where $\sigma_\infty^\mx$ and $\sigma_\infty^\mn$  are the maximal and minimal principal {\it in situ} stresses, respectively, $\be_1$ and $\be_2$ are  unit vectors along the coordinate axis. We assume that the prestress $\btau^0$ satisfy the same boundary condition: $\btau^0\langle\bn\rangle|_{\Gamma_R} = \bsigma_\infty$.
 
The line $y=0$ is divided into the part $\Gamma_f = \{ -L_\ell(t) \leqslant x \leqslant L_r(t),\; y=0 \}$ occupied
by the fracture and the remaining part $\Gamma_s = \{ -R<x<-L_\ell(t),\; y=0\}\bigcup \{ L_r(t)<x<R,\; y=0\}$. Outside the fracture on  $\Gamma_s$ the symmetry conditions (see.~\cite{ShelBaikGol2014}) are satisfied:
\begin{equation}\label{eq:symmBC}
		\Gamma_s:\quad \pd{u}{y}=0, \quad v=0, \quad \pd{p}{y}=0.
\end{equation}

With $p_f(t,x)$ standing for the fluid pressure inside the fracture, the force balance over the fracture's wall yields
	\begin{equation}\label{eq:fracBC}
		\Gamma_f:\quad p=p_f, \quad \bn\cdot\btau\langle\bn\rangle=-p_f + \sigma_{coh}, \quad \bbs\cdot\btau\langle\bn\rangle=0,
	\end{equation}
	where $\sigma_{coh}$ is a cohesive stress implementing energy dissipation and material failure during fracturing~\cite{Golovin_Baykin_2016_Pore}.
Here we neglect the tangential stress due to the fluid friction on the	fracture's walls in comparison to the normal stress. The fluid flow in the fracture is governed by the mass conservation law complemented with the Poiseuille formula:	
	\begin{equation}\label{eq:lubric_equation}
		\pd{w}{t}+\pd{(wq)}{x}=-q_l,\quad w\equiv v|_{y=0},\quad q=-\frac{(2w)^2}{12\eta_f}\pd{p_f}{x}.
	\end{equation}
	Here $w$ is a half of the fracture's aperture, $q$ is the fluid velocity in $x$-direction. No fluid lag is assumed at the fracture tip.
	
	The leak-off velocity $q_l$ is given by the Darcy law as
	\begin{equation}\label{leakoff}
		q_l=-\left.\frac{k_r}{\eta_r}\pd{p}{y}\right|_{y=0}.
	\end{equation}
	
	The resulting equation governing the flow inside the fracture reads
	\begin{equation}\label{eq:mass_conserve_lubricBC}
		\pd{w}{t}=\pd{}{x}\left(\frac{w^3}{3\eta_f}\pd{p_f}{x}\right)+\left.\frac{k_r}{\eta_r}\pd{p}{y}\right|_{y=0}.
	\end{equation}
	
	The flow rate (per unit height) injected into the fracture upper half-plane is calculated as 
	\begin{equation}\label{eq:flowrateBC}
		Q(t) = \frac{Q_v(t)}{2H}= -\left.\frac{w^3}{3\eta_f}\pd{p_f}{x}\right|_{x=0+} + \left.\frac{w^3}{3\eta_f}\pd{p_f}{x}\right|_{x=0-},
	\end{equation}
	where the division by 2 shows that the total flow rate is equally distributed between the symmetric fracture's parts for positive and negative $y$ and $Q_v(t)$ denotes the volumetric flow rate injected into the well. Equation \eqref{eq:mass_conserve_lubricBC} is referred to as the lubrication theory equation~\cite{Adachi}.

The initial data at some moment $t^0$ reads:
	\begin{equation}\label{eq:init_cond}
		\bu|_{t=t^0} = \bu^0(x,y), \quad p|_{t=t^0} = p^0(x,y),\quad L_i|_{t=t^0} = L^0_i,\; i=\ell,r.
	\end{equation}
	
	For computational reasons, it is convenient to homogenise the	conditions over the outer boundary $\Gamma_R$. It can be done by considering the stresses inside the reservoir relative to the prestress state $\btau^0$ and taking  $p_\infty$ as the reference pressure. Similar to \cite{ShelBaikGol2014}, the following new sought functions are introduced:
	\begin{equation}\label{eq:tau_tild}
		\tilde{\bu} = \bu -  \varkappa \bx, \quad \varkappa = \dfrac{\alpha p_\infty}{2(\lambda+\mu)}, \quad \tilde{\btau} = \btau -\btau^0, \quad \tilde{p} = p - p_\infty.
	\end{equation}	
	Substituting \eqref{eq:tau_tild} into equations \eqref{eq:model} and taking into account boundary conditions \eqref{eq:GammaRBC}, \eqref{eq:symmBC}, \eqref{eq:fracBC},  \eqref{eq:mass_conserve_lubricBC}, \eqref{eq:flowrateBC}, we obtain the following problem
	\begin{flalign}
		\displaystyle \Omega: & \quad \dv{\tilde{\btau}}=-\dv{\btau^0}, \quad \tilde{\btau} =  \lambda \dv{\tilde{\bu}}\id+2 \,\mu\, \epss{(\tilde{\bu})}-\alpha \tilde{p} \id, & \label{eq:tild_formul_equilib} \\[4mm]
		\displaystyle \Omega: & \quad S_\varepsilon\pd{\tilde{p}}{t} =  \dv{\Bigl( \dfrac{k_r}{\eta_r} \nabla \tilde{p} - \alpha \pd{\tilde{\bu}}{t} \Bigr)},& \label{eq:tild_formul_filtr}  \\[4mm]
		\displaystyle \Gamma_R: & \quad \tilde{p}=0, \quad  \tilde{\btau}\langle\bn\rangle = 0, & \label{eq:tild_formul_gamma_R}  \\[4mm]
		\displaystyle \Gamma_s: & \quad \tilde{u_y}=0, \quad \tilde{v} = 0, \quad \tilde{p_y} = 0, & \label{eq:tild_formul_gamma_s}  \\[4mm]
		\displaystyle \Gamma_f: & \quad  \bn \cdot \tilde{\btau}\langle\bn\rangle = -(\tilde{p}+p_\infty) + \sigma_{coh} - \bn \cdot \btau^0\langle\bn\rangle ,\quad \bbs \cdot \tilde{\btau}\langle\bn\rangle = 0, & \label{eq:tild_formul_gamma_f_force} \\[4mm]
		\displaystyle \Gamma_f: & \quad \pd{\tilde{v}}{t}=\pd{}{x}\left(\frac{\tilde{v}^3}{3\eta_f}\pd{\tilde{p}}{x}\right)+\frac{k_r}{\eta_r}\pd{\tilde{p}}{y}; & \\[4mm]
		&\quad -\frac{\tilde{v}^3}{3\eta_f}\pd{\tilde{p}}{x}\Bigr|_{y=0,x=0+} + \frac{\tilde{v}^3}{3\eta_f}\pd{\tilde{p}}{x}\Bigr|_{y=0,x=0-} 
		= Q(t). & \label{eq:tild_formul_gamma_f_lubric} 		
	\end{flalign}
	In what follows, we work with the new sought functions skipping the tilde for simplicity of notations.	

	In this paper we do not discuss the details of the numerical algorithm for computation of the fracture propagation, nor the implementation of the rock failure criteria, referring the reader to paper \cite{Golovin_Baykin_2016_Pore} where the numerical algorithm is described in details and verified against known solutions. 

\section{The formal definition of the backstress}\label{sec:backstress}
It is commonly accepted that the pressure of pore fluid creates the so-called backstress that impedes the deformation of the poroelastic medium. In paper \cite{Golovin_Baykin_2016_Pore} this property was formulated as the thumb rule 'pore pressure stiffens the rock'. In this section we give an analytical expression for the backstress and demonstrate its action on a simple exact solution describing one-dimensional compression of a layer of a poroelastic medium. In the forthcoming sections this estimation is used for the numerical calculation of the backstress and interpretation of the simulation results for fracture propagation in heterogeneous medium. 

Let us represent the displacement vector $\bu$ and the stress tensor $\btau$ as
\[\bu = \bu^r+\bu^p,\quad \btau = \bsigma^r+\bsigma^p-\alpha p\id,\]
where
\[\bsigma^k = \lambda\dv\bu^k\id+2\mu\epss(\bu^k), \quad k=r,p. \]
Given the pore pressure $p$ defined by a solution of the problem \eqref{eq:tild_formul_equilib}--\eqref{eq:tild_formul_gamma_f_lubric} we demand tensor $\bsigma^p$ to satisfy the following boundary value problem: 
\begin{equation}\label{eq:sigmab}
\begin{array}{ll}
&\dv\bsigma^p = \alpha\nabla p, \\[2mm]
\Gamma_R: &\bsigma^p\langle\bn\rangle = 0,  \\[2mm]
\Gamma_s\cup \Gamma_f: & \partial u^p/\partial y=0, \quad v^p = 0.
\end{array}
\end{equation}
Tensor $\bsigma^p$ describes the stress caused purely by the ``volume force'' $\alpha\nabla p$ that corresponds to the action of the pore pressure and does not contribute to the fracture openning. Hence, the remaining tensor $\bsigma^r$ satisfies 
	\begin{equation}\label{eq:sigmareq}
		\begin{array}{ll}
			&\dv\bsigma^r = -\dv\btau^0, \\[2mm]
			\Gamma_R: & \bsigma^r\langle\bn\rangle = 0,  \\[2mm]
			\Gamma_s: & \partial u^r/\partial y=0, \quad v^r = 0, \\[2mm]
			\Gamma_f: & \bn \cdot \bsigma^r\langle\bn\rangle = -(p+p_\infty) - \bn \cdot \btau^0\langle\bn\rangle +\sigma_{coh} +\underline{\alpha p - \bn \cdot \bsigma^p\langle\bn\rangle},\\[2mm]
			&\bbs \cdot \bsigma^r\langle\bn\rangle = 0.
		\end{array}
	\end{equation}
Tensor $\bsigma^r$ describes the deformation of the elastic skeleton of the reservoir, subjected only to the boundary load over the fracture and to the prestress $\btau^0$. The underlined terms in the expression for the normal boundary stress over the fracture $\Gamma_f$ in \eqref{eq:sigmareq} correspond to the additional pressure on fracture's walls due to the presence of the pore pressure. We treat these two terms as the analytical expression of the backstress. 

\begin{dfn} 
	Given the pore pressure $p$ as a part of the solution of the boundary value problem \eqref{eq:tild_formul_equilib}--\eqref{eq:tild_formul_gamma_f_lubric} and tensor $\bsigma^p$ satisfying equations \eqref{eq:sigmab} the following stress acting over the fracture wall
\begin{equation}\label{eq:backstress}
\sigma^b = (\alpha p - \bn \cdot \bsigma^p\langle\bn\rangle)|_{\Gamma_f}
\end{equation}
is referred to as the backstress.
\end{dfn}

\begin{figure}[t]
		\centering
			\includegraphics[width=0.2\linewidth]{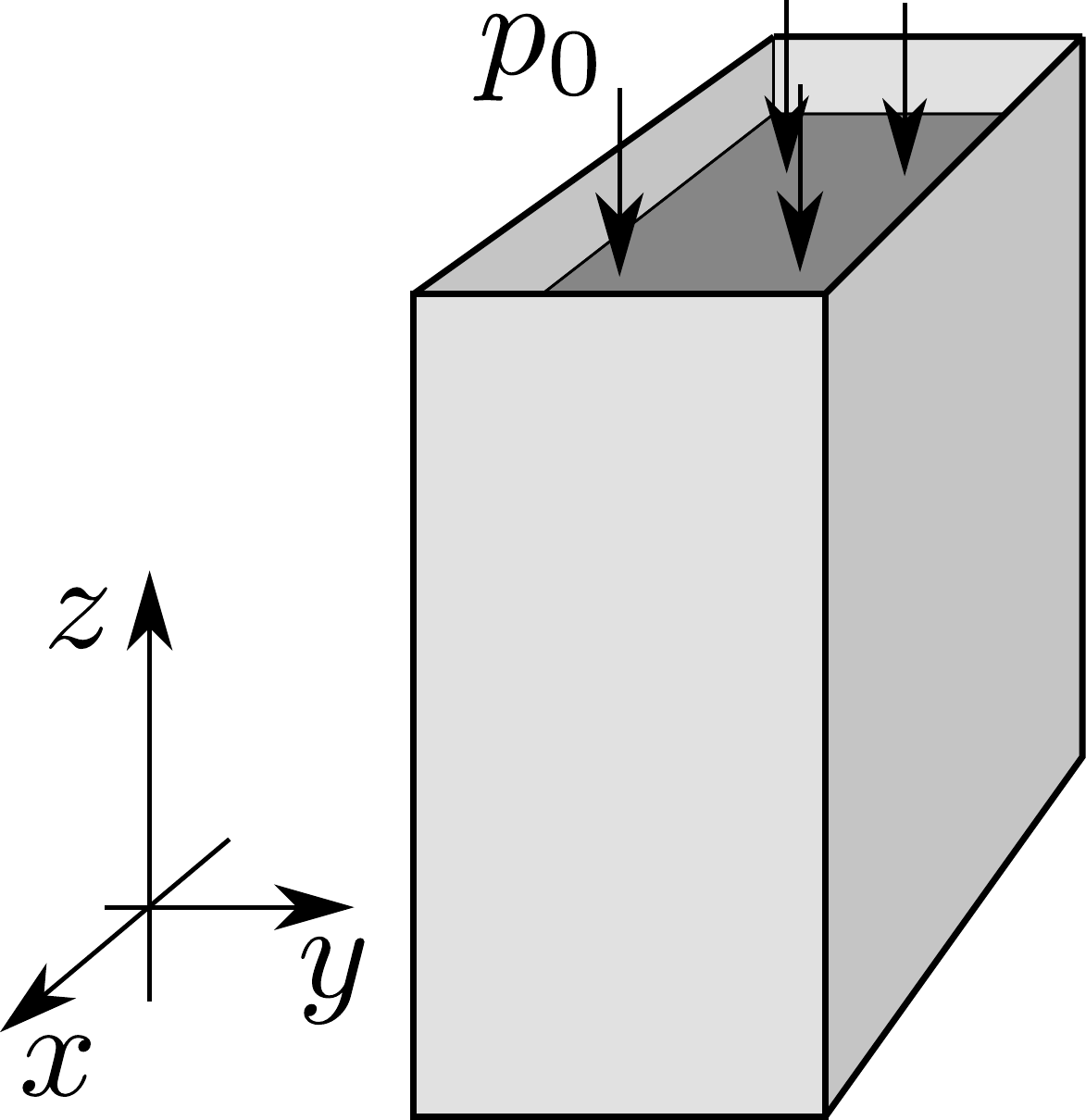}
		\caption{Deformation of a poroelastic specimen placed in a rectangular rigid impermeable box and loaded on top by the fluid pressure $p_0$.}
		\label{fig:BackStressProblem}
	\end{figure}

Let us demonstrate this definition on an exact solution of a model problem. The statement of the problem reads as follows. Consider a poroelastic specimen in the shape of a parallelepiped located between two pairs of parallel rigid impermeable walls. One end of the specimen rests against a rigid impermeable bottom. The opposite end is covered with fluid under pressure $p_0(t)$. It is required to find the stress and the pore pressure within the specimen (see Figure \ref{fig:BackStressProblem}).

For the solution of the stated problem we place the Cartesian coordinate system as shown in Figure \ref{fig:BackStressProblem} such that the bottom and the top ends of the specimen correspond to $z=0$ and $z=L$ respectively. We assume that deformation of the specimen takes place only in the vertical direction $z$: $\bu = \bigl(0,0,w(t,z)\bigr)^T$. At that, the strain tensor has only one non-zero component:
\[\varepsilon_{33}=\pd{w}{z},\quad \varepsilon_{ij}=0, \; ij\ne 33.\]
Under this assumption the equilibrium equation $\dv\btau = 0$ with the boundary condition 
\[\btau\langle\bn\rangle = -p_0(t)\bn \quad \mbox{at } z=L\]
give 
\begin{equation}\label{eq:stress_solution}
(\lambda+2\mu)\pd{w}{z}-\alpha p=-p_0(t).
\end{equation}
Substitution of the derivative $\partial w/\partial z$ from equation \eqref{eq:stress_solution} into the equation for the pore pressure (the second equation of \eqref{eq:model}) yields
\begin{equation}\label{eq:pressure_solution}
\left(S_\varepsilon+\frac{\alpha^2}{\lambda+2\mu}\right)\pd{p}{t}=\frac{k}{\mu}\frac{\partial^2 p}{\partial z^2}+\frac{\alpha}{\lambda+2\mu}p_0'(t).
\end{equation}
Equations \eqref{eq:stress_solution}, \eqref{eq:pressure_solution} are to be solved with the following boundary conditions:
\begin{equation}\label{eq:boundary_pressure}
w|_{z=0}=0,\quad p|_{z=L}=p_0(t), \quad \left.\pd{p}{z}\right|_{z=0}=0.
\end{equation}
Integration of equation \eqref{eq:stress_solution} gives the expression for the displacement of the upper boundary of the specimen:
\begin{equation}\label{eq:displacement_solution}
w=-\frac{L}{\lambda+2\mu}\left(p_0(t)-\frac{\alpha}{L}\int_{0}^{L}p(t,s) ds\right).
\end{equation}
The first term in \eqref{eq:displacement_solution} corresponds to the exact solution of the problem for the pure elastic medium (no pore pressure, $\alpha = 0$), whereas the second term describes the action of the backstress. Hence, in this solution the backstress can be naturally defined as 
\begin{equation}\label{eq:BackstressNaive}
\sigma^b = \frac{\alpha}{L}\int_0^L p(t,s)ds.
\end{equation}
Note, that the backstress acts against the compressive stress $p_0(t)$. 

Let us now obtain the same formula by using definition \eqref{eq:backstress}. For the displacement vector under the same assumption of one-dimensional deformation we have $\bu^p=\bigl(0,0,w^p(t,z)\bigr)^T$. The problem for the stress tensor $\bsigma^p$ is reduced to 
\begin{equation}\label{eq:wb}
(\lambda+2\mu)\frac{\partial^2 w^p}{\partial z^2}=\alpha p,\quad w^p|_{z=0}=w^p|_{z=L}=0.
\end{equation}
Here $p(t,z)$ is the pressure given by the solution of equation \eqref{eq:pressure_solution} with the boundary conditions \eqref{eq:boundary_pressure}. The solution of problem \eqref{eq:wb} reads
\[w^p=\frac{\alpha}{\lambda+2\mu}\left(\int_0^zp(t,s)ds-\frac{z}{L}\int_0^L p(t,s)ds\right).\]
Hence,
\[\bn\cdot\bsigma^p\langle\bn\rangle|_{z=L}=(\lambda+2\mu)\pd{w^p}{z}|_{z=L}=\alpha p_0(t)-\frac{\alpha}{L}\int_0^L p(t,s)ds.\]
According to the definition \eqref{eq:backstress}, we obtain (cf. \eqref{eq:BackstressNaive}):
\[\sigma^b = (\alpha p - \bn \cdot \bsigma^p\langle\bn\rangle)|_{z=L} = \frac{\alpha}{L}\int_0^L p(t,s)ds.\]
Later in this paper we use the definition \eqref{eq:backstress} for the numerical calculation of the backstress.
\section{Heterogeneous reservoir}
	
		\begin{table}[t]
			\renewcommand{\arraystretch}{1.2} 
			\begin{center}	
				\caption {Input parameters for the reference verification case} \label{tab:sim_parameters_verification}				\begin{tabular}{|@{\hspace{2em}}l@{\hspace{2em}}|@{\hspace{2em}}c@{\hspace{2em}}|}
					
					\hline	
					{\bf Parameter} & {\bf Value} \\
					\cline{1-2}
					Domain size, $R$  &  $150$ m\\
					Max. right tip position, $L_r^\mx$  &  $73$ m\\
					Max. left tip position, $L_\ell^\mx$  &  $40$ m\\
					Young's modulus, $E$  &  $17$ GPa\\
					Poisson's ratio, $\nu$ &  $0.2$ \\
					Energy release rate, $G_c$ & 120 Pa$\cdot$m \\
					Critical cohesive stress, $\sigma_c$ & 1.25 MPa \\
					Initial porosity, $\phi_0$ & 0.2 \\
					Reservoir permeability, $k_r$ &  $10^{-14}$ m$^2$\\ 
					Biot's coefficient, $\alpha$ &  $0.75$ \\
					Min. far-field stress, $\sigma_\infty^\mn$ &  $10$ MPa \\
					Reservoir pressure, $p_\infty$ &  $0$ MPa \\
					Reservoir fluid viscosity, $\eta_r$ &  $10^{-3}$ Pa$\cdot$s\\
					Fracturing fluid viscosity, $\eta_f$ &  $10^{-1}$ Pa$\cdot$s \\
					Injection rate per unit height, $2\,Q$ &  $10^{-3}$ m$^2/$s\\
					\hline
					
				\end{tabular}
				
			\end{center}
		\end{table}

Let us now use the described model to study the influence of the inhomogeneity of the reservoir's physical properties and of the variable in space confining stresses on the fracture propagation.  In the numerical experiments we assume that the prestress has the form
	\[
		\btau^0 = \left(
			\begin{array}{cc}
				-\sigma_\infty^\mx & 0 \\
				0 & -\sigma_\infty^\mn
			\end{array}\right),
	\]
	where $\sigma_\infty^\mx$ is constant and  $\sigma_\infty^\mn(x)$ is a piecewise constant. Following~\cite{Golovin_Baykin_2016_Pore}, it can be shown that $\sigma_\infty^\mx$ does not affect the final problem solution.
	
The physical parameters used in simulations are listed in Table~\ref{tab:sim_parameters_verification}, unless otherwise noted.  

\subsection{Closure stress contrast } \label{sec:closure_stress_contrast}
\begin{figure}[t]
	\centering
	\includegraphics[width=0.7\textwidth]{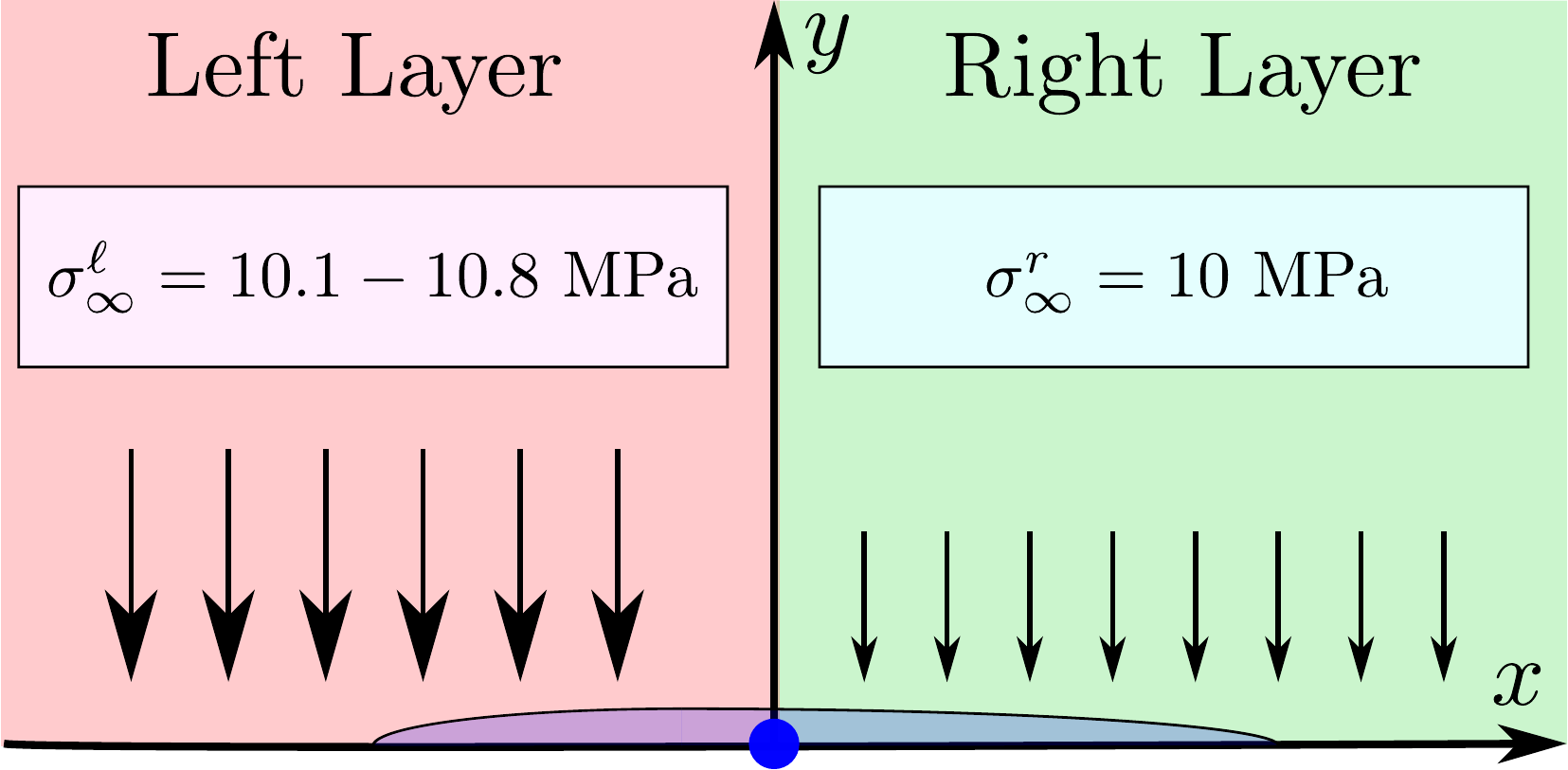}
	\caption{Reservoir with the different formation closure stress for negative (left) and positive (right) values of $x$.}
	\label{fig:two_layers_scheme_half}
\end{figure}

First, we demonstrate how the non-constant minimal {\it in situ} confining stress influences on the symmetry of fracture's wings. We divide  the reservoir to two layers: the right one for $x>0$ and the left one for $x\le0$, and suppose that the minimal confining stress $\sigma^\mn_\infty$ is different in each layer due to some geological or physical reasons (see Figure \ref{fig:two_layers_scheme_half}): 
	\begin{equation}\label{eq:layered_stress}
		\sigma_\infty^\mn(x) = 
		\left\{
			\begin{array}{ll}
				\sigma^\ell_\infty, & x \leqslant 0, \\[1ex]
				\sigma^r_\infty, & x > 0.\\[1ex]
			\end{array}
		\right.
	\end{equation}

The goal is to check if the fracture would propagate non-symmetrically in view of the non-symmetry of the outer stress. Keeping the confining stress $\sigma^r_\infty=10$ MPa in the right layer fixed, we perform computations for $\sigma^\ell_\infty = 10.1, 10.2,10.4,10.6$, and $10.8$ MPa.	

\begin{figure}[t]
		\begin{minipage}[t]{0.49\linewidth} 
			
			\includegraphics[width=1\linewidth]{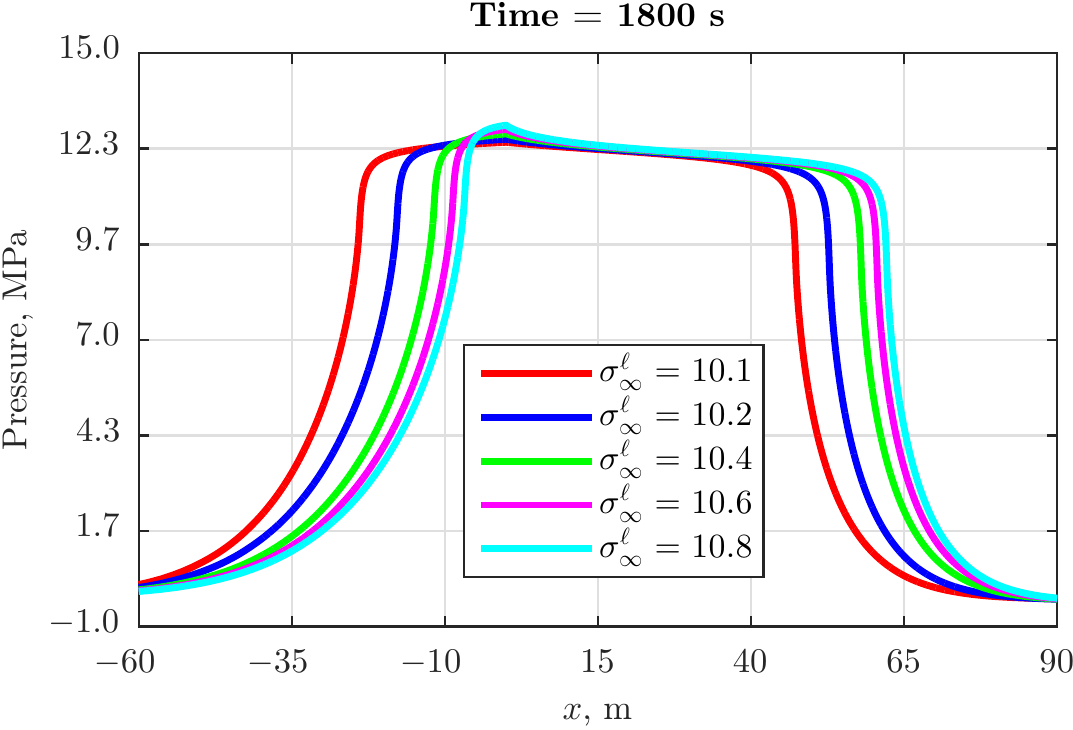}
			\centering {\bf(a)}
	
		\end{minipage}
		\hspace{0.02\linewidth}
		\begin{minipage}[t]{0.49\linewidth}
			
			\includegraphics[width=1\linewidth]{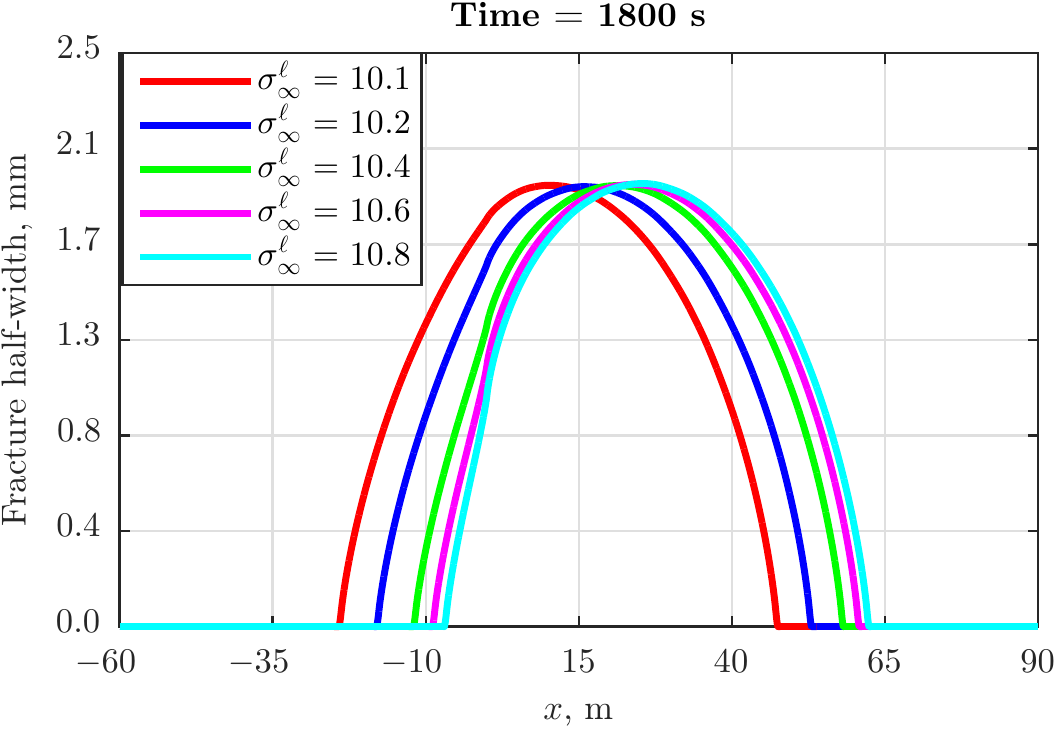}
			\centering{\bf(b)}
			
		\end{minipage}
	
		\caption{Pressure (a) and fracture half-width (b) along the fracture at $t=1800$ s for different values of the closure stress $\sigma^l_\infty$.}
		\label{fig:pressureAndFracWidthProfileOnStressTimestep950}
	
	\end{figure}

The result of the simulations is shown in Figure \ref{fig:pressureAndFracWidthProfileOnStressTimestep950}, where the pressure and the fracture half-width distribution along the fracture at $t=1800$ s are presented. One can see that increase of the contrast of the confining stresses $\sigma^l_\infty/\sigma^r_\infty>1$ leads to the development of the non-symmetry of the fractureэы wings. The left wing with the higher value of the confining stress almost stops its propagation even for the small, of the order of 1\%, contrast.  

\begin{figure}[t]
		\begin{minipage}[t]{0.49\linewidth} 
			
			\includegraphics[width=1\linewidth]{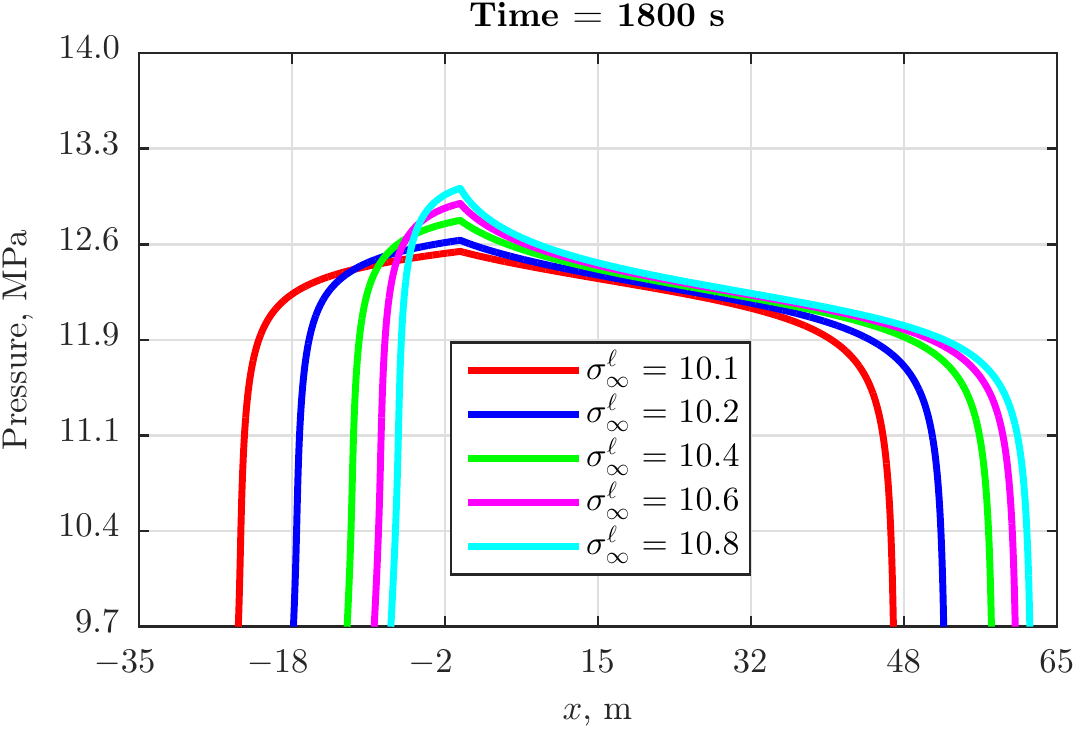}
				\caption{Pressure along the fracture at $t=1800$ s for different values of the closure stress $\sigma_\infty^\ell$  (zoom of Figure \ref{fig:pressureAndFracWidthProfileOnStressTimestep950} (a)) }
				\label{fig:pressureProfileOnStressZOOMEDTimestep950}	
			
		\end{minipage}
		\hspace{0.02\linewidth}
		\begin{minipage}[t]{0.49\linewidth}
			
			\includegraphics[width=1\linewidth]{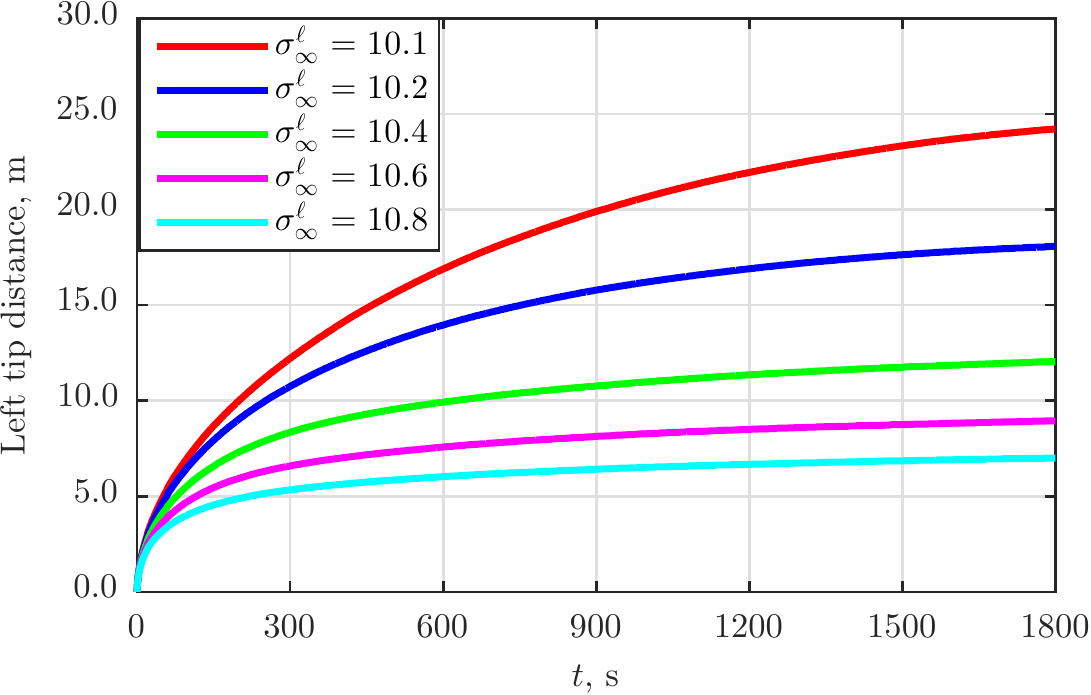}
				\caption{The distance of the left tip $L_\ell$ from the injection point vs. time. Left fracture's tip almost stops for larger contrast of the confining stresses}
				\label{fig:LeftTipDistanceOnStress}	
		\end{minipage}
	\end{figure}

The explanation of the non-symmetry is straightforward: the fracture aperture (see Figure~\ref{fig:pressureAndFracWidthProfileOnStressTimestep950}~(b)) on the right-hand side is wider due to the smaller confining stress. Higher aperture results in the smaller hydrodynamic resistance to the flow. In combination with  the  higher pressure gradient right to the injection point (see Figure \ref{fig:pressureProfileOnStressZOOMEDTimestep950}) this generates an enhanced flow directed to the right wing of the fracture. Therefore, the left fracture's tip almost stops due to the lack of the fluid inflow (see Figure \ref{fig:LeftTipDistanceOnStress}).

This numerical experiments evidences that even a small contrast of confining stresses causes the significant non-symmetry of the fracture wings, especially for large fractures. The difference of the confining stresses acting over the two wings of the fracture can be caused not only by the difference of the {\it in situ} stresses, which might be observed as non-natural from the geological point of view, but also by the non-uniformity of physical properties of the reservoir due to the difference in the backstress caused by the pore pressure. This effect is demonstrated in the next section.

	\subsection{Permeability Contrast}
It was demonstrated in paper~\cite{Golovin_Baykin_2016_Pore} and in Section \ref{sec:backstress} above that the pore pressure plays the significant role in the stresses redistribution near the fracture. Action of the pore pressure can be treated as an additional stress applied to the fracture's wall due to the pore pressure within the reservoir.

The numerical experiment of the previous section demonstrates that the difference in the confining stress causes the non-symmetry of the fracture even for small contrast of the stresses acting on fracture's wings. In this section we demonstrate how this effect can be caused by the difference of the backstresses due to the difference of the rock permeability in the neighbouring layers.
	
\begin{figure}[t]
		\centering
		\includegraphics[width=0.7\textwidth]{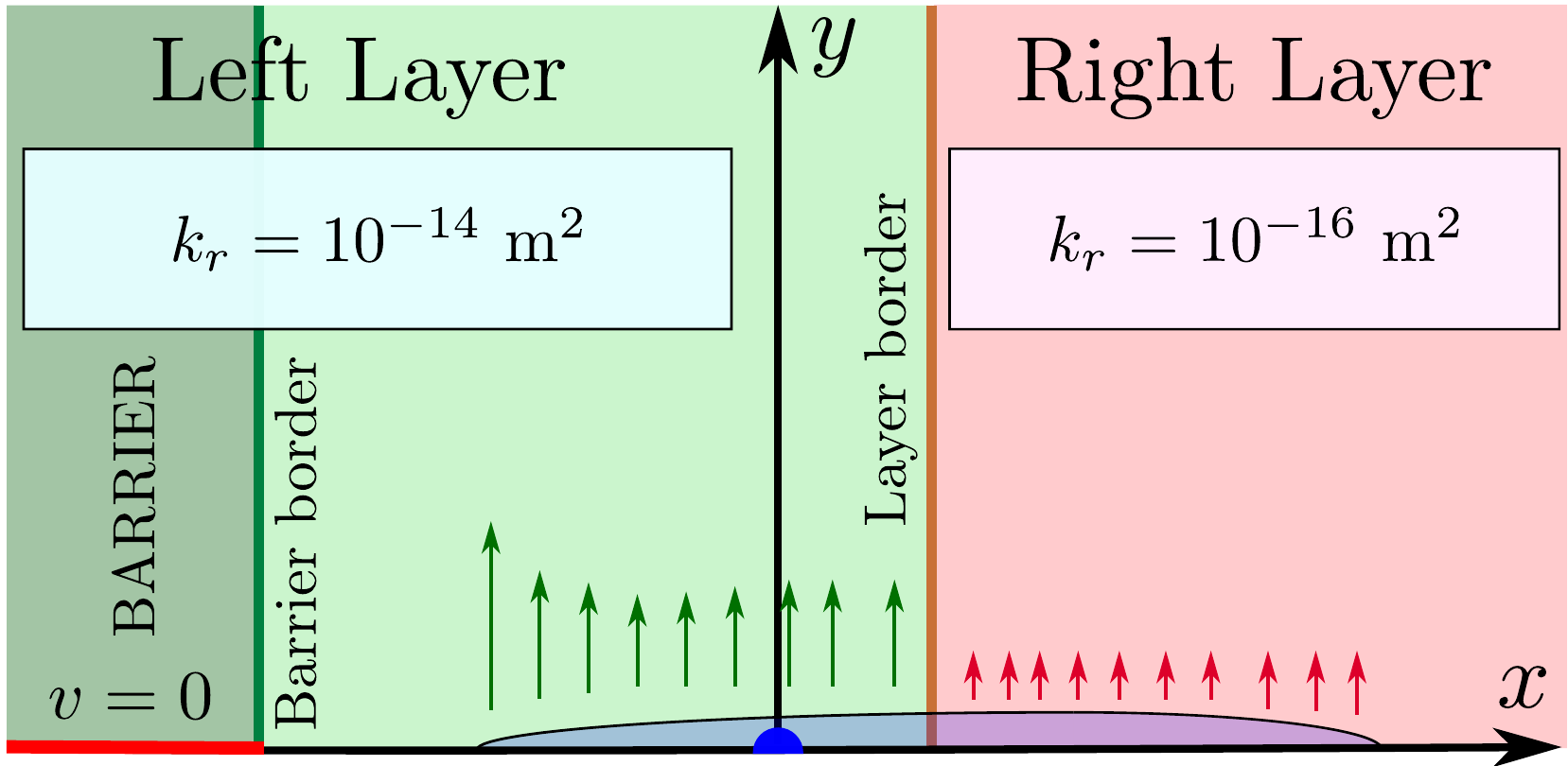}
		\caption{Reservoir with high-permeable (left) and low-permeable (right)  layers. The barrier is located in high-permeable layer, preventing the further fracture propagation. Arrows sketch the velocity of the filtration of fluid to the reservoir.}
		\label{fig:two_layers_scheme_permeability_half}
	\end{figure}

For the numerical experiment we construct the reservoir with two layers (see Figure \ref{fig:two_layers_scheme_permeability_half}) that have the following permeabilities: 
	\begin{equation}\label{eq:layered_perm}
	k_r(x) = 
	\left\{
	\begin{array}{ll}
	10^{-14} \text{ m}^2, & x \leqslant x^\ast, \\[1ex]
	10^{-16} \text{ m}^2, & x > x^\ast.\\[1ex]
	\end{array}
	\right.
	\end{equation}  
Here $x=x^\ast>0$ denotes the border between the layers, shown as the brown solid line parallel to the $y$-axis in Figure \ref{fig:two_layers_scheme_permeability_half}. Thus, the leakoff from the fracture into the reservoir is higher in the left layer. For the reasons that will be clarified later, we put the barrier for the fracture development at $x=-L_\ell^\mx$  as shown in Figure \ref{fig:two_layers_scheme_permeability_half}. Presence of the barrier implies that in the numerical experiment the fracture cannot penetrate the barrier (it models either a high-toughness material or a layer with a very high confining stress).

	\begin{figure}[t]
		\begin{minipage}[t]{0.49\linewidth} 
			
            \ifthenelse{\equal{\theisElectronicVersion}{1}}{
                \animategraphics[autoplay,width=\textwidth,loop,poster = 55] {1500}{HeterPermPropRight/pressure_anim/pressureProfileTimestep}{1}{56}    
            }{
                \includegraphics[width=1\textwidth]{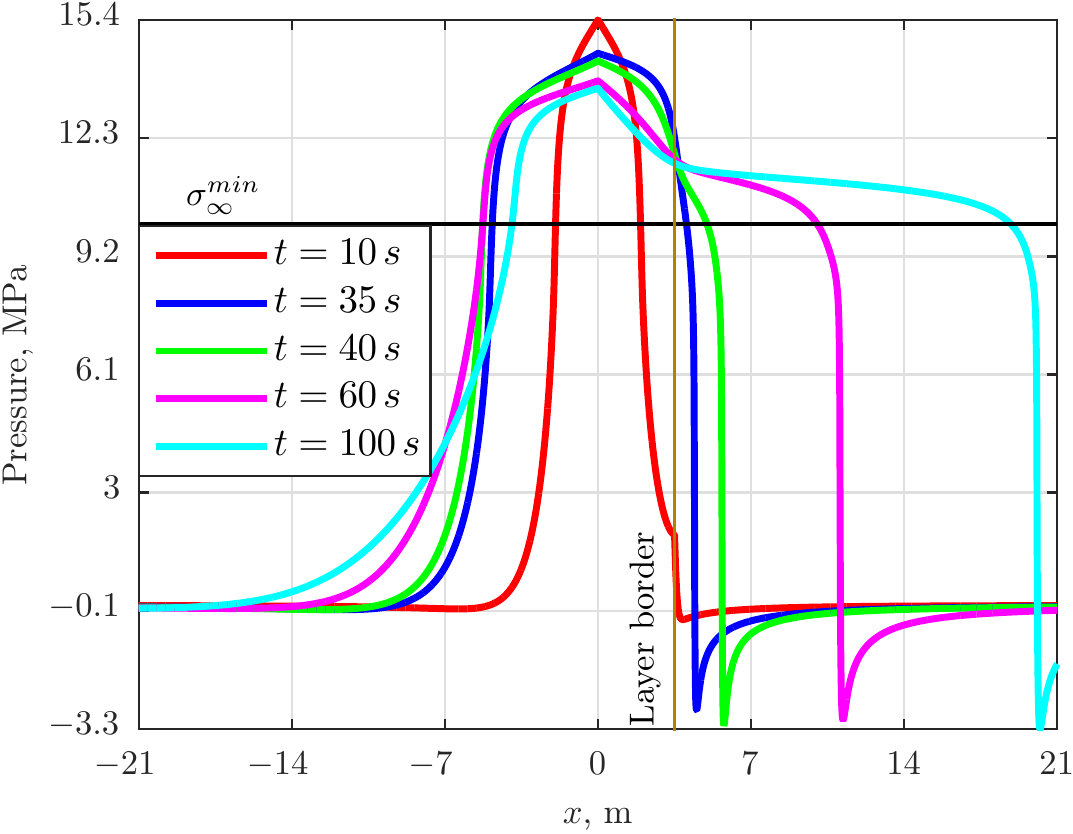}
            }
			\centering {\bf(a)}
		\end{minipage}
		\hspace{0.02\linewidth}
		\begin{minipage}[t]{0.49\linewidth}
	
            \ifthenelse{\equal{\theisElectronicVersion}{1}}{
                \animategraphics[autoplay,width=\textwidth,loop,poster = 55] {1500}{HeterPermPropRight/width_anim/widthProfileTimestep}{1}{56}
            }{
                \includegraphics[width=1\textwidth]{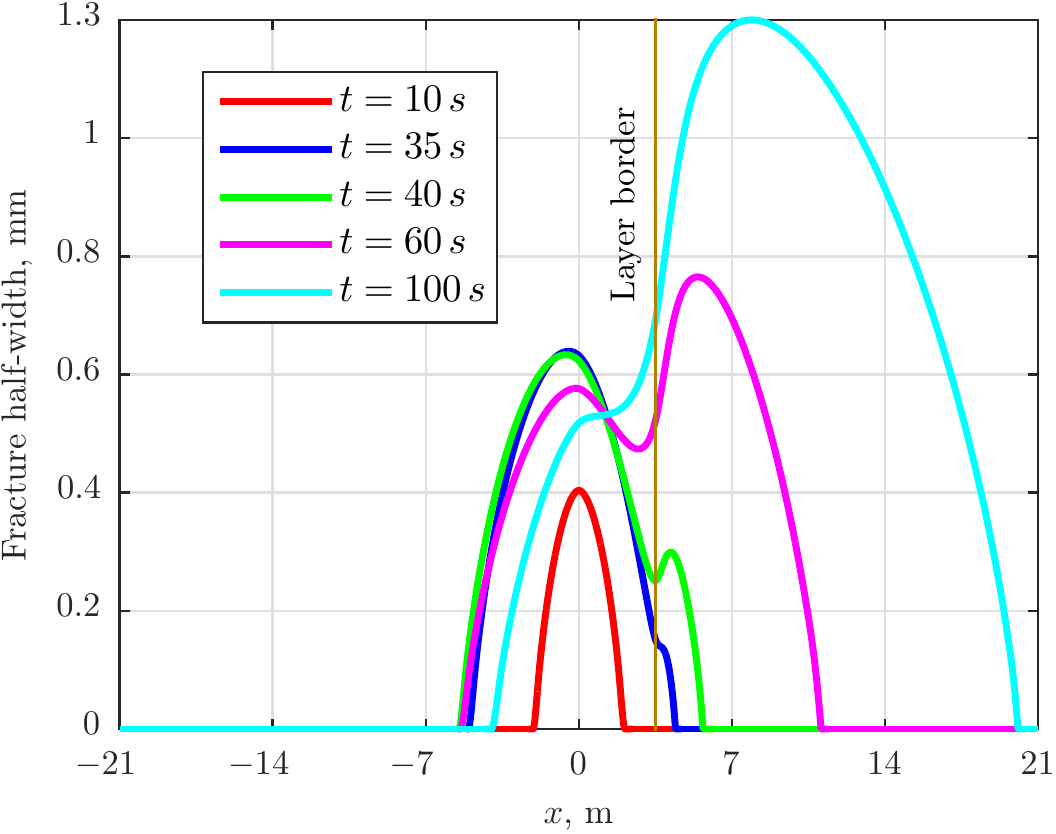}
            }
			\centering{\bf(b)}
			
		\end{minipage}
		
		\caption{Pressure (a) and fracture half-width (b) along the fracture at different time for reservoir with two zones with  high permeability  contrast. The zones border (vertical brown line) is located at $x=3.5$ m }
		\label{fig:pressureAndFracWidthProfilePerm2RightDifferentTime}
		
	\end{figure}

Our goal is to study the dynamics of fracture propagation depending on the location $x=x^\ast$ of the border between the layers relative to the injection point.  

In the first experiment we place the layers' border close to the injection point at $x^*=3.5$ m.  Figure~\ref{fig:pressureAndFracWidthProfilePerm2RightDifferentTime} shows the snapshots of the pressure distribution and the fracture aperture along the fracture path at increasing time moments. One can see that the fracture propagates symmetrically until it reaches the layers' border at $t\approx 35$ s. Then it rapidly breaks into the low-permeable (right) layer and propagates only to the right layer such that the left fracture's tip stops. 

\begin{figure}[t]
    
    \ifthenelse{\equal{\theisElectronicVersion}{1}}{
        \centering
        \animategraphics[autoplay,width=0.7\textwidth,loop,poster = 50] {1500}{HeterPermPropRight/backstress_anim/pressureProfileTimestep}{1}{100}
    }{
        \begin{minipage}[t]{0.49\linewidth} 
                      
            \includegraphics[width=1\textwidth]{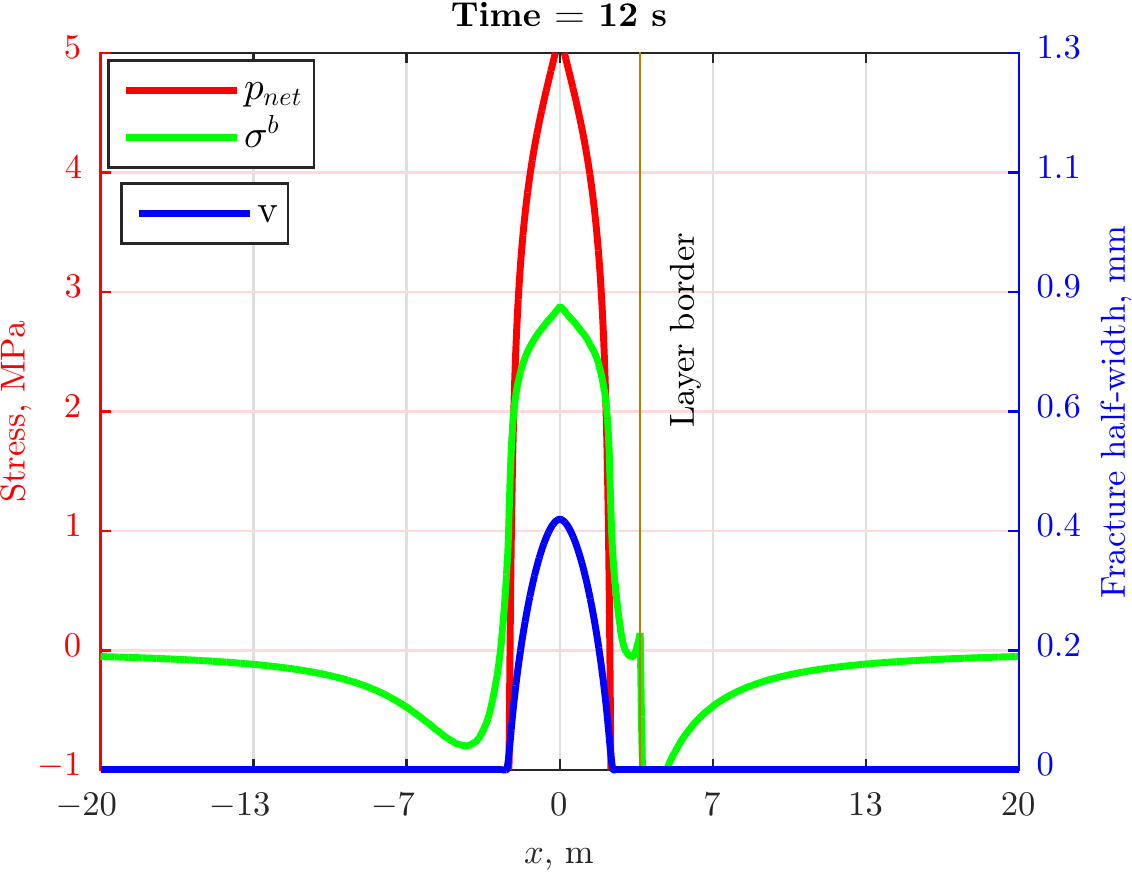}
        
            \centering {\bf(a)}   
             
        \end{minipage}
        \hspace{0.02\linewidth}
        \begin{minipage}[t]{0.49\linewidth} 
            
            \includegraphics[width=1\textwidth]{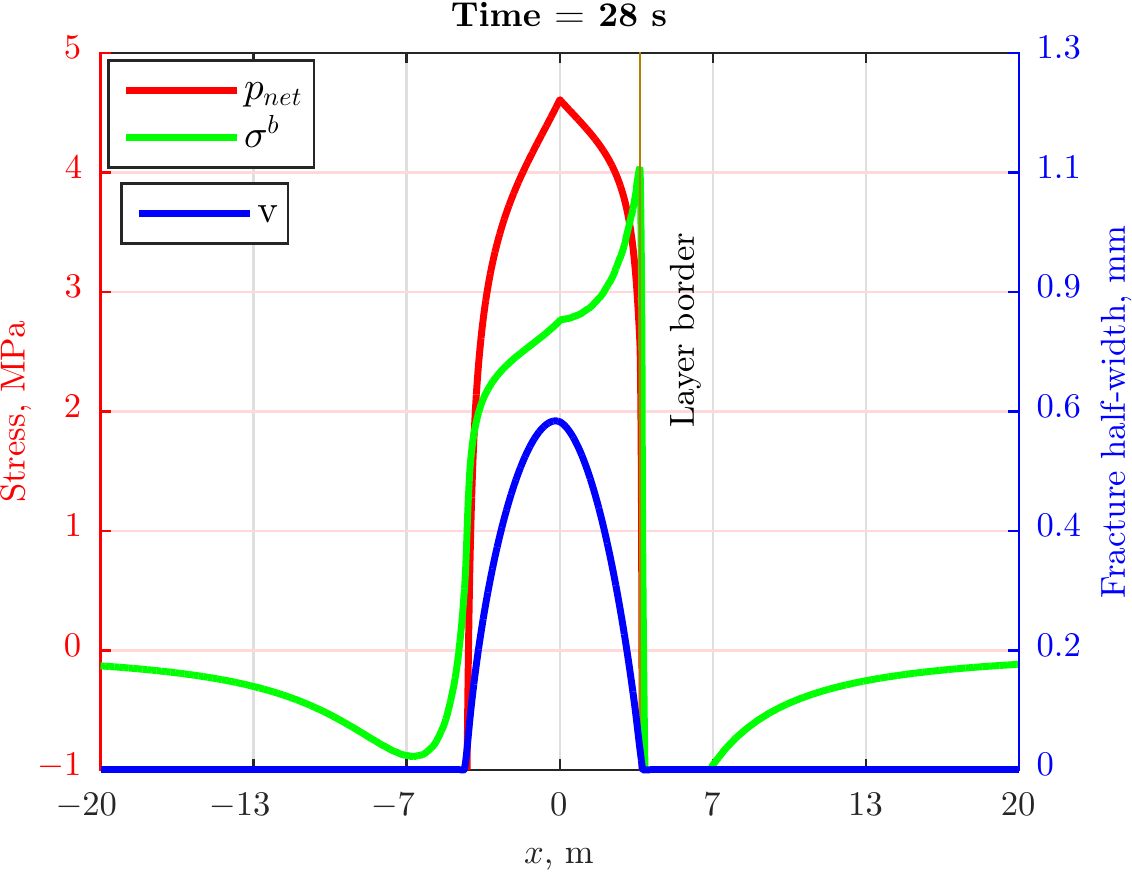}
        
            \centering {\bf(b)}
        \end{minipage}
        
        \vspace{2ex}      
          
        \begin{minipage}[t]{0.49\linewidth} 
            
            \includegraphics[width=1\textwidth]{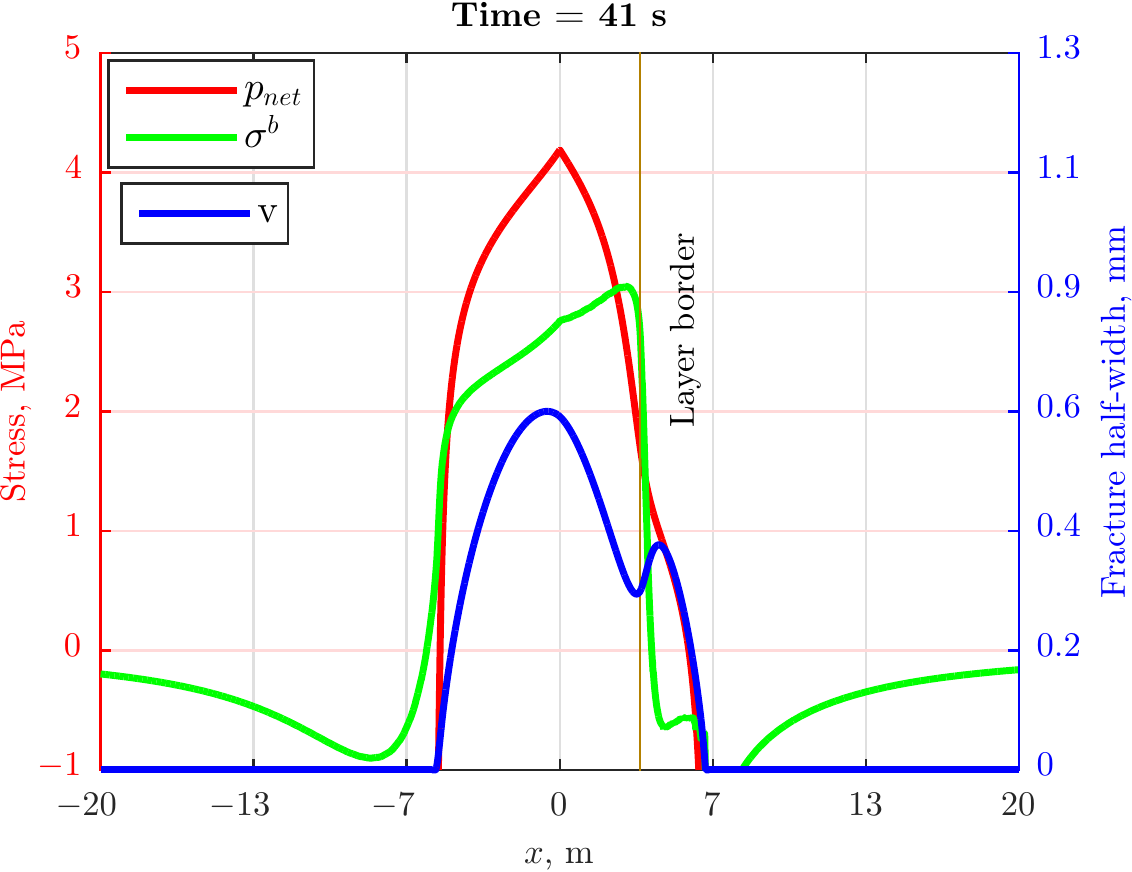}
        
            \centering {\bf(c)}   
        
        \end{minipage}
        \hspace{0.02\linewidth}
        \begin{minipage}[t]{0.49\linewidth} 
            
            \includegraphics[width=1\textwidth]{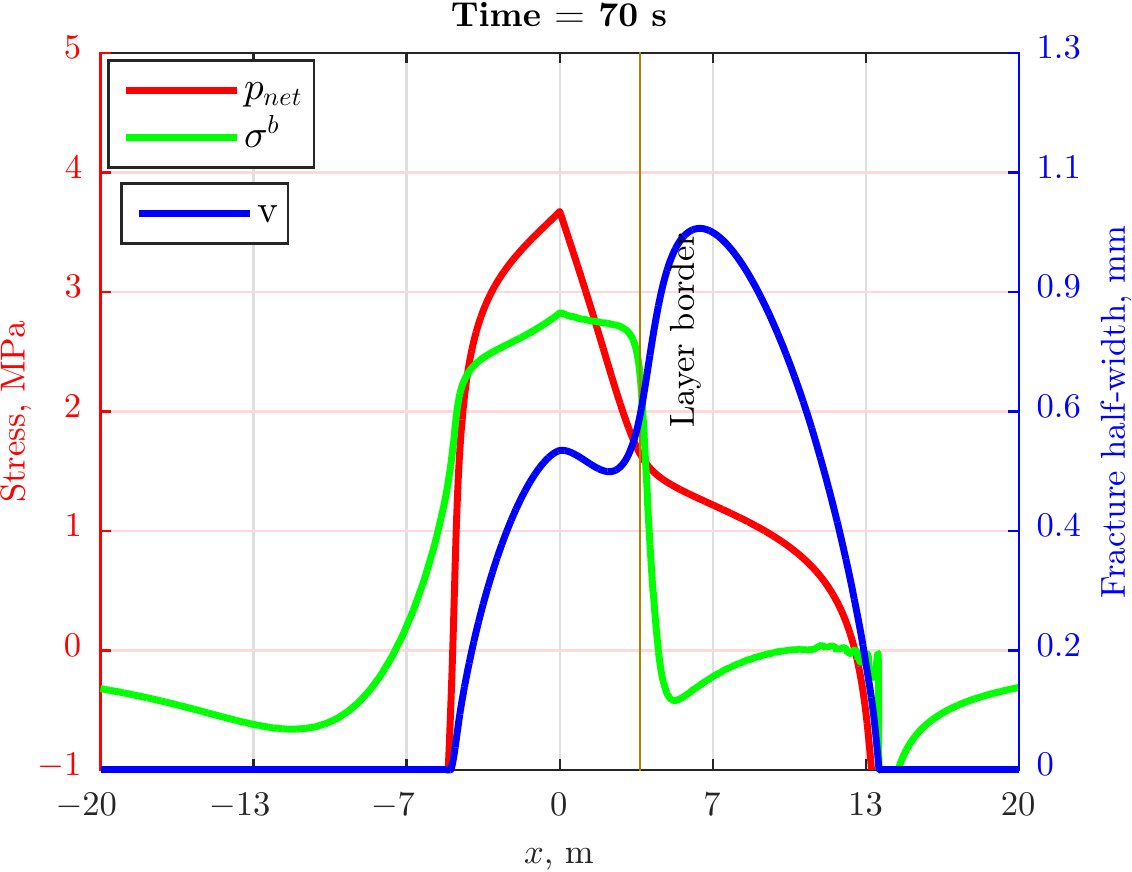}
        
            \centering {\bf(d)}
            
        \end{minipage}
    }
    
    \caption{Dynamics of the net pressure $p_\mathrm{net}$, the backstress $\sigma^b$ and the fracture's half-width in the numerical experiment where the border between layers with different permeabilities is located at $x^\ast = 3.5$ m. Live in electronic version   }\label{fig:backstressPropRight}

\end{figure}

This behaviour can be explained by the combination of the two factors: formation of the higher backstress in the left layer due to the higher fluid filtration, and by the lower fluid loss in the right layer due to the lower leakoff. The dynamics of the net pressure 
\[p_\mathrm{net}=(p+p_\infty) + \bn \cdot \btau^0\langle\bn\rangle,\]
the backstress and the fracture's half-width supporting this conclusion is demonstrated in Figure~\ref{fig:backstressPropRight}.

Remarkable, that in this experiment the fracture's aperture becomes non-convex. Another interesting feature is the negative pressure that is developed in front of the fast propagating right fracture's tip. This effect is caused by the rapid deformation of the medium during the fracture propagation and by the influence of the deformation velocity of the material to the fluid filtration (see the Second thumb rule in \cite{Golovin_Baykin_2016_Pore}). This implies that the porous fluid is sucked to the area in front of the fracture's tip from the neighbouring region as the fracture avalanches to the right.

\begin{figure}[t]
	\begin{minipage}[t]{0.49\linewidth} 
		\ifthenelse{\equal{\theisElectronicVersion}{1}}{
			\animategraphics[autoplay,width=\textwidth,loop,poster = 100] {1500}{HeterPermPropLeft/pressure_anim/pressureProfileTimestep}{1}{166}
		}{
			\includegraphics[width=1\textwidth]{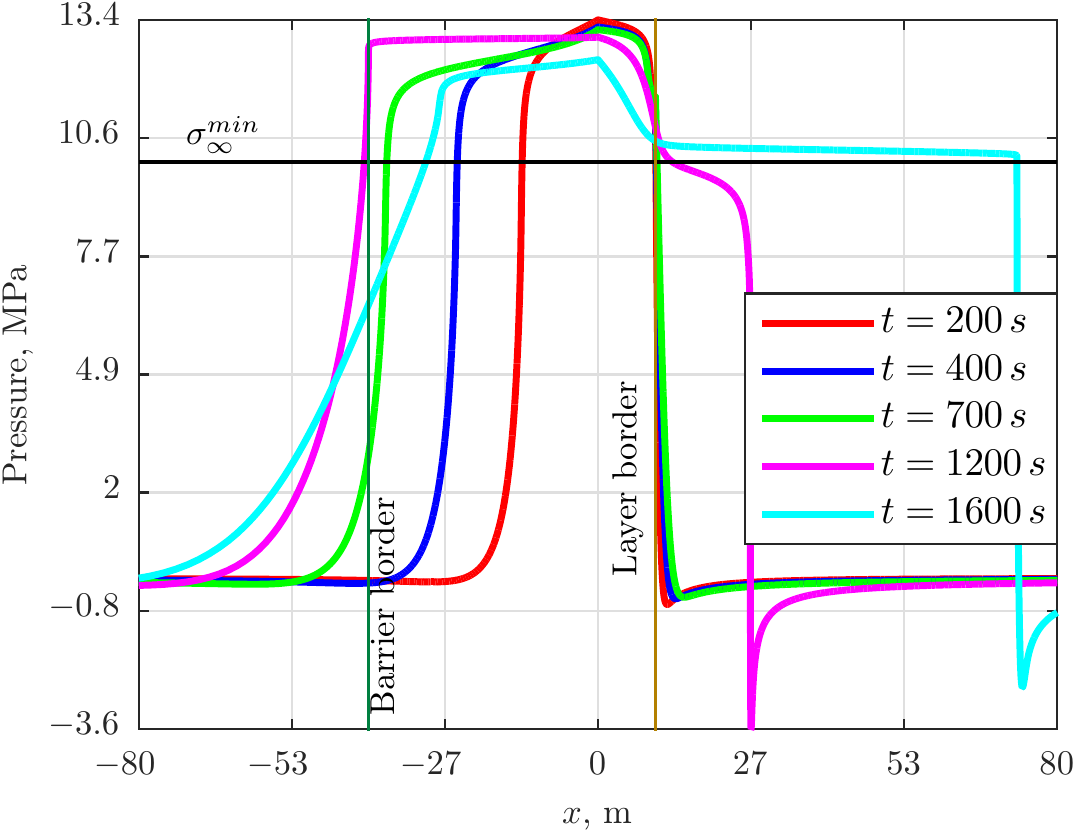}
		}	
		\centering {\bf(a)}
		
	\end{minipage}
	\hspace{0.02\linewidth}
	\begin{minipage}[t]{0.49\linewidth}
		\ifthenelse{\equal{\theisElectronicVersion}{1}}{
			\animategraphics[autoplay,width=\textwidth,loop,poster = 100] {1500}{HeterPermPropLeft/width_anim/widthProfileTimestep}{1}{166}
		}{
			\includegraphics[width=1\textwidth]{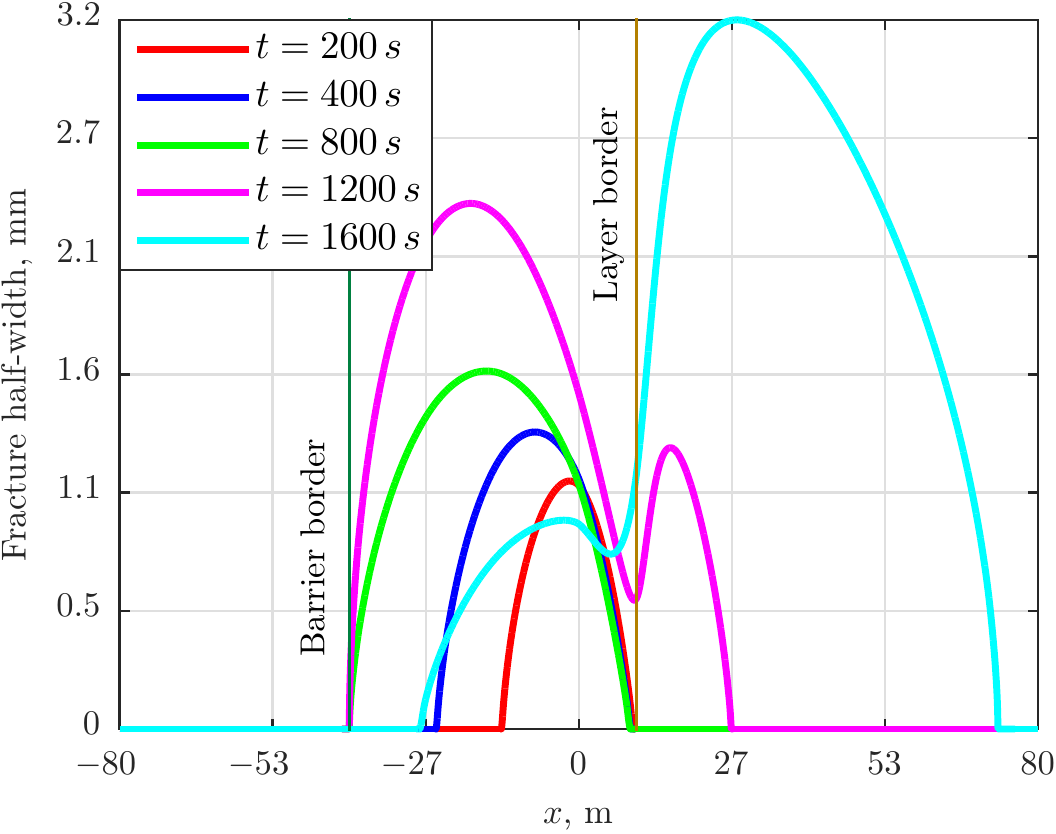}
		}			
		\centering{\bf(b)}
		
	\end{minipage}
	
	\caption{Pressure (a) and fracture half-width (b) along the fracture at different time for reservoir with two zones with  high permeability  contrast. The zones border (vertical brown line) is located at $x=10$ m }
	\label{fig:pressureAndFracWidthProfilePerm2LeftDifferentTime}
	
\end{figure}

Another scenario of the fracture propagation is realized if we shift the layers' border to $x^*=10$ m (see Figure \ref{fig:pressureAndFracWidthProfilePerm2LeftDifferentTime}). In this case the fracture propagates in both directions until the right tip almost reaches the  border at $t\approx200$ s. Then, the right tip of the fracture stops and the fracture propagates only to the left until the left tip reaches the barrier at $t\approx 1200$ s. For the time period $1200\le t\le 1600$, both tips of the fracture remain in the same positions while the aperture of the fracture increases to accommodate fluid injected at the wellbore. Finally, at $t\approx 1600$ s the fracture breaks into the low-permeable layer at $x>x^\ast$ and rapidly propagates to the right such that the left wing of the fracture shuts-in. Continuation of the fluid injection leads to the unlimited growth of the right wing of the fracture while the left wing remains small of a fixed length.

\begin{figure}[t]
	
	\ifthenelse{\equal{\theisElectronicVersion}{1}}{
		\centering
		\animategraphics[autoplay,width=0.7\textwidth,loop,poster = 26] {1500}{HeterPermPropLeft/backstress_anim/pressureProfileTimestep}{1}{166}
	}{
		\begin{minipage}[t]{0.49\linewidth} 
			
			\includegraphics[width=1\textwidth]{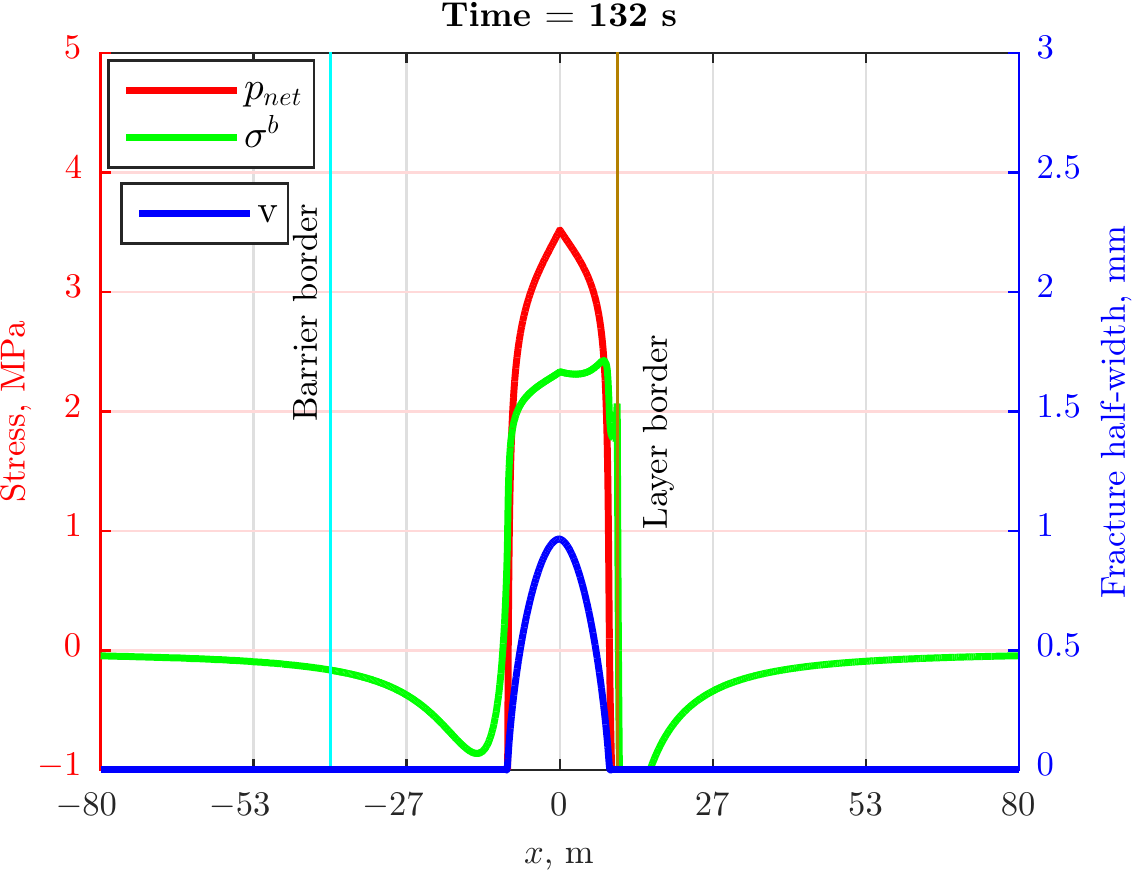}
			
			\centering {\bf(a)}   
			
		\end{minipage}
		\hspace{0.02\linewidth}
		\begin{minipage}[t]{0.49\linewidth} 
			
			\includegraphics[width=1\textwidth]{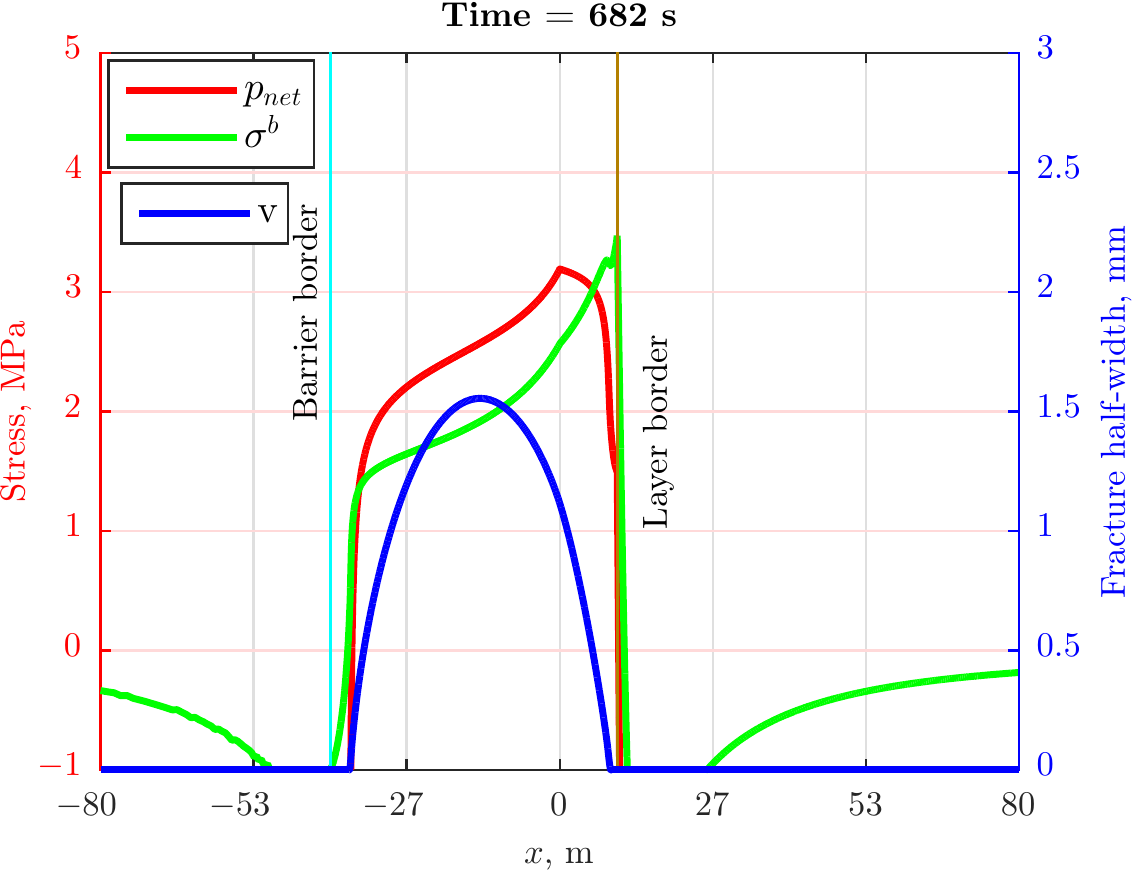}
			
			\centering {\bf(b)}
		\end{minipage}
		
		\vspace{2ex}      
		
		\begin{minipage}[t]{0.49\linewidth} 
			
			\includegraphics[width=1\textwidth]{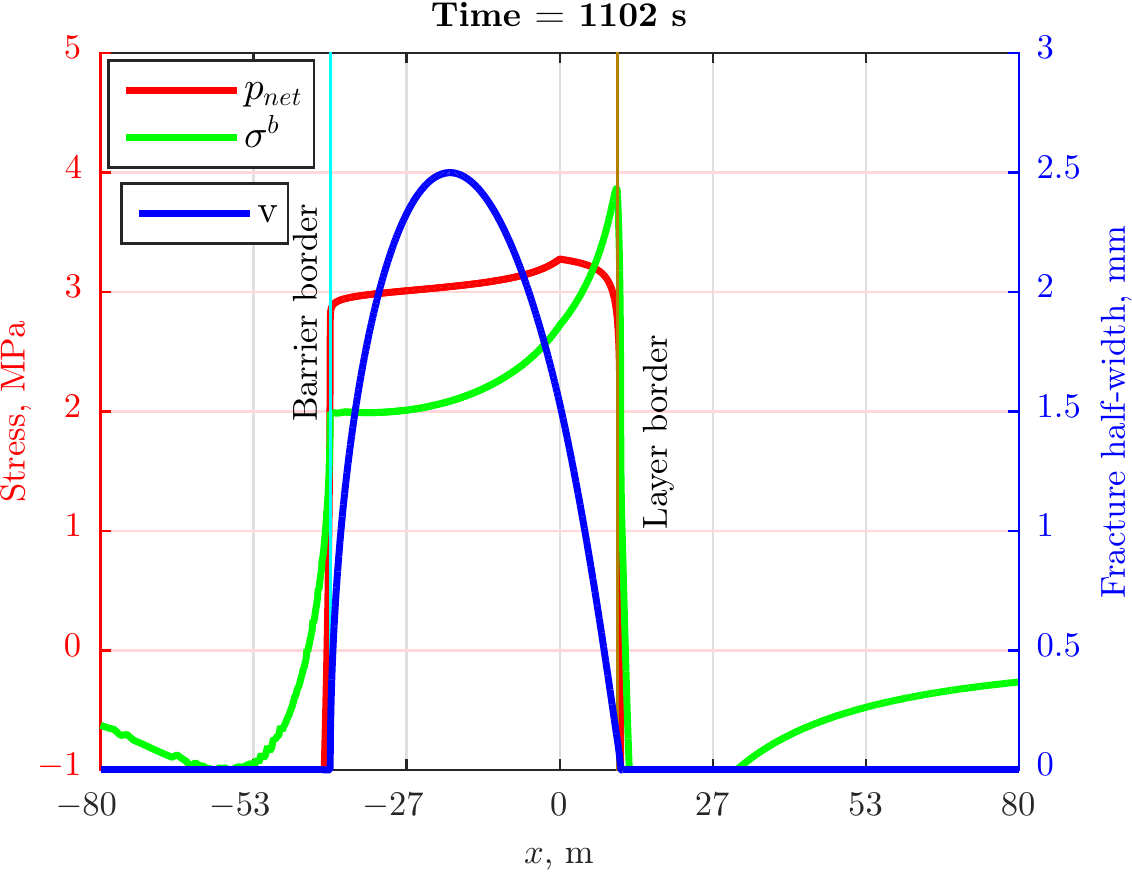}
			
			\centering {\bf(c)}   
			
		\end{minipage}
		\hspace{0.02\linewidth}
		\begin{minipage}[t]{0.49\linewidth} 
			
			\includegraphics[width=1\textwidth]{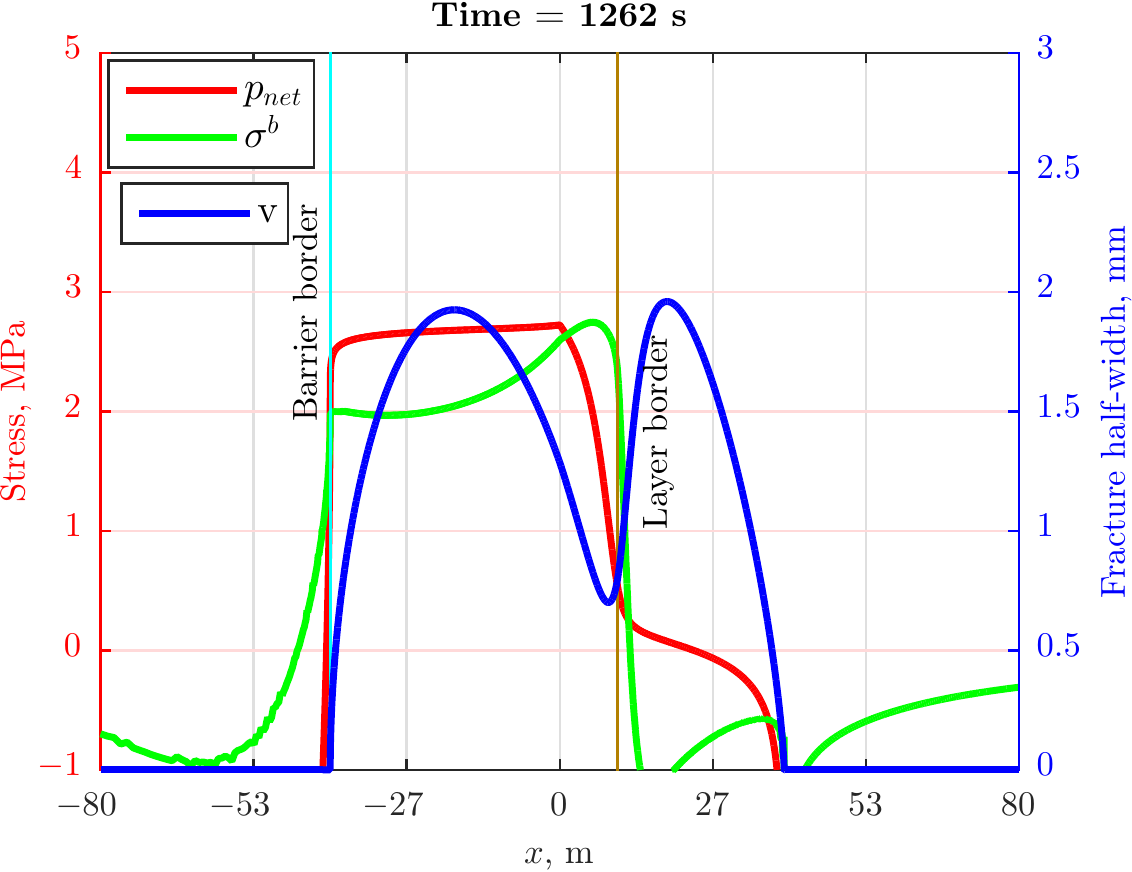}
			
			\centering {\bf(d)}
			
		\end{minipage}
	}
	
	\caption{Dynamics of the net pressure $p_\mathrm{net}$, the backstress $\sigma^b$ and the fracture's half-width in the numerical experiment where the border between layers with different permeabilities is located at $x^\ast = 10$ m. 
	}\label{fig:backstressPropLeft}
	
\end{figure}

Interpretation of such a  behaviour can be done in terms of the interplay between the backstress and the fluid filtration. The process starts in the high-permeable layer for $x<x^\ast$ where the rapid filtration of fluid to the reservoir creates a noticeable backstress. As the right tip of the fracture approaches the border with the low-permeable reservoir, the filtration front sets against the border and the filtrating fluid propagates to the sides of the fracture at the same time increasing the backstress to the right on the injection point  (Figure \ref{fig:backstressPropLeft}). This effectively creates the contrast of confining stresses that results in the non-symmetry of the fracture and in its further extension to the area of the lower backstress. 

As the fracture reaches the barrier, it cannot propagate further to the left, hence the injected fluid increases the volume of the fracture and enlarges the area of the fluid filtration near the fracture. This process continues until the stress on the fracture walls to the left of the wellbore exceeds the stress on the walls in the right part. After that the fracture breaks into the low-permeable formation where it propagates rapidly due to the lower fluid leakoff and the smaller value of the backstress.

\subsection{Fracture propagation in an uncoupled medium}
\begin{figure}[t]
		\begin{minipage}[t]{0.49\linewidth} 
			\ifthenelse{\equal{\theisElectronicVersion}{1}}{
                \animategraphics[autoplay,width=\textwidth,loop,poster = 100] {1500}{HeterPermPropRightZeroAlpha/pressure_anim/pressureProfileTimestep}{1}{166}
            }{
                \includegraphics[width=1\textwidth]{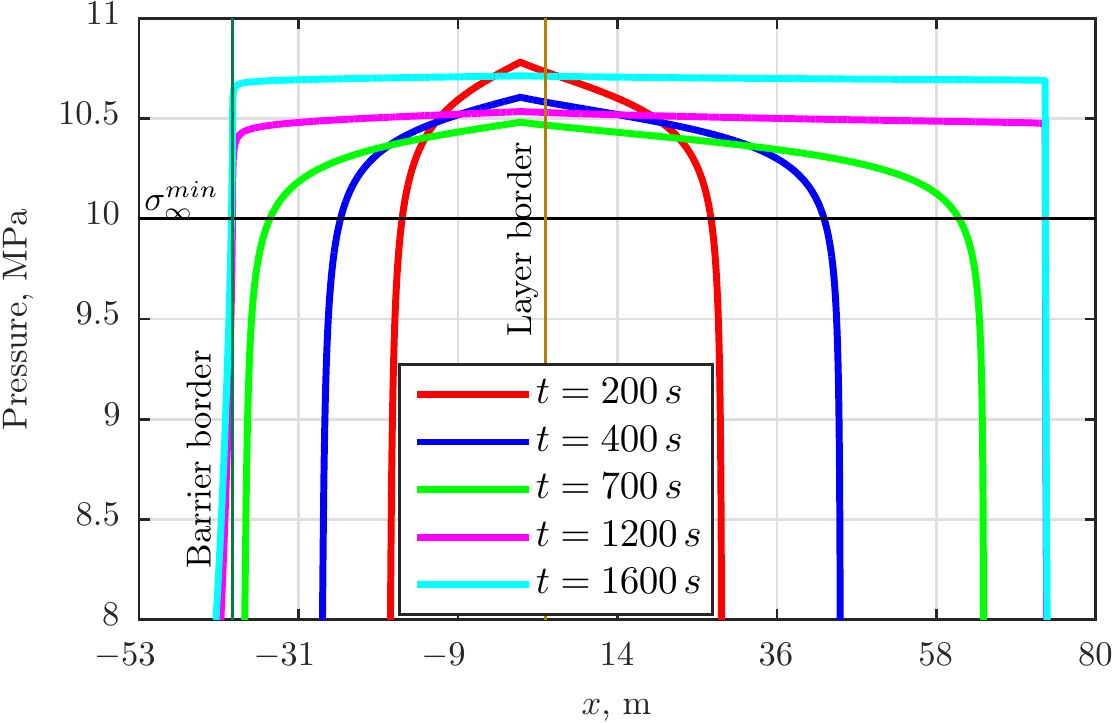}
            }
			\centering {\bf(a)}
			
		\end{minipage}
		\hspace{0.02\linewidth}
		\begin{minipage}[t]{0.49\linewidth}
			\ifthenelse{\equal{\theisElectronicVersion}{1}}{
                \animategraphics[autoplay,width=\textwidth,loop,poster = 100] {1500}{HeterPermPropRightZeroAlpha/width_anim/widthProfileTimestep}{1}{166}
            }{
                \includegraphics[width=1\textwidth]{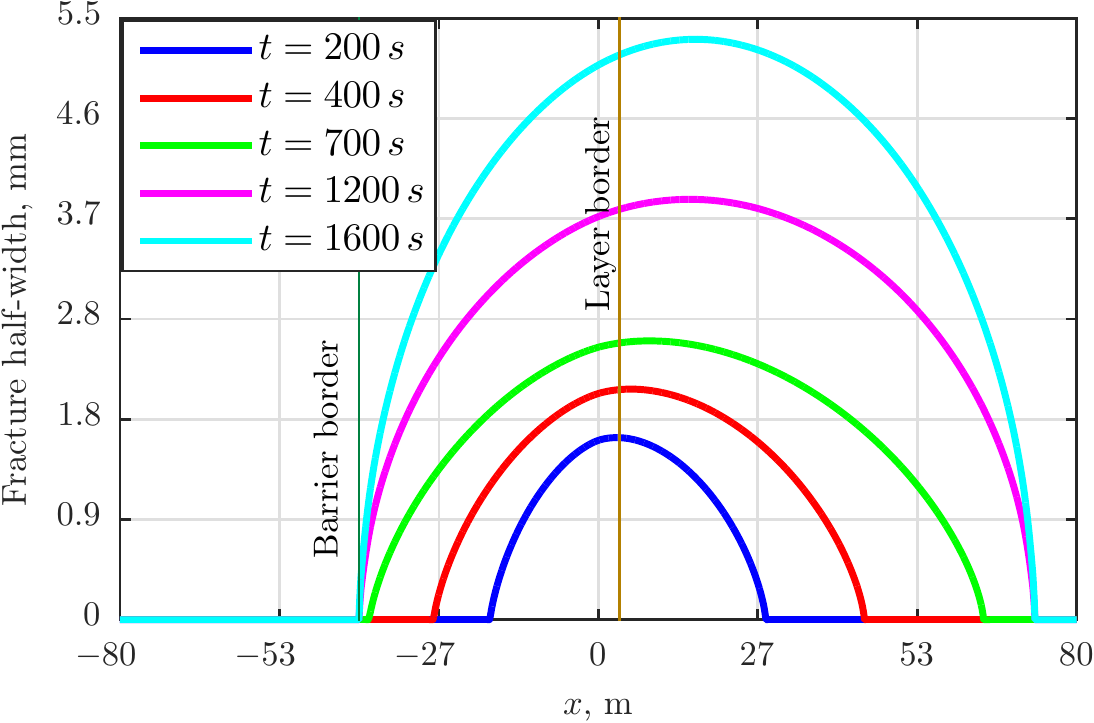}
            }
			\centering{\bf(b)}
			
		\end{minipage}
		
		\caption{Pressure (a) and fracture half-width (b) along the fracture at different time for reservoir with two zones with  high permeability  contrast. The zones border (vertical brown line) is located at $x=3.5$ m. The Biot's coefficient is $\alpha=0$. }
		\label{fig:pressureAndFracWidthProfileAlphaZeroDifferentTime}
		
	\end{figure}

In order to support the interpretation of the fracture's behaviour in terms of the action of the backstress, we repeat the same numerical experiment as in previous section for the fully uncoupled medium where the stresses and the fluid filtration do not interact. This medium corresponds to zero Biot's coefficient: $\alpha=0$. In this case the pore fluid does not influences the elastic stresses in the reservoir. The calculations show that under this conditions the process of fracture propagation is not affected by the difference in rock permeabilities as it was in the previous example. Indeed, as follows from Figure \ref{fig:pressureAndFracWidthProfileAlphaZeroDifferentTime}, fracture propagates in both layers with some preference to the low-permeable right layer due to the smaller fluid loss because of the lower permeability.  
	
\section{Conclusion and discussion}
In this paper we used the numerical model developed in \cite{Golovin_Baykin_2016_Pore} for the simulation of the hydraulic fracture propagation in an inhomogeneous poroelastic medium. We demonstrate that the in case of the non-constant confining {\it in situ} stresses the fracture extends non-symmetrically, i.e. the wing in the layer with the higher confining stress stops whereas the wing in the layer with the lower stress extends. The same effect of the fracture non-symmetry can be reached in case of inhomogeneous physical properties of the reservoir due to the formation of the backstress caused by the pressure of pore fluid. We gave an analytical definition of the backstress and demonstrated the reliability of the definition on a simple example of one-dimensional compression of a poroelastic specimen due to the fluid pressure applied to one side of the specimen. 

We performed numerical experiment for fracture propagation in a composite medium such that two neighbouring semi-spaces have different permeabilities. The fracture is originated in the higher-permeable part and propagates towards the lower-permeable region. As one of two fracture's tips approaches the border between the regions, the fracture either breaks into the low-permeable region or stops depending on the distance between the injection point and the border. In the case of the small distance the fracture avalanches into the low-permeable part such that another fracture's tip in the higher-permeable region stops or even this fracture's wing shuts-in. This behaviour is explained by the lower fluid loss due to the leakoff in the low-permeable reservoir and, as a consequence, by the lower backstress in the low-permeable reservoir. On contrary, if the distance between the injection point and the border between the two layers is large enough, the backstress formation becomes non-symmetrical as the fracture approaches the border. In this case the backstress contrast between the two fracture's wings becomes high and causes the fracture propagation only into the low-permeable reservoir. 

The described effects of non-symmertical fracture propagation due to the inhomogeneity of the resrvoir's permeability are explained solely by the action of the backstress. To support this conclusion we performed the same calculation for the case of uncoupled medium where the pore pressure and fluid filtration are not linked with the stress in the reservoir (i.e., the Biot's number is zero: $\alpha=0$). In this case the backstress is zero and we obtained  almost symmetrical picture of fracture propagation with a small non-symmetry of the two wings explained by the difference of the leakoffs in the two regions. 

The non-symmetry of fracture propagation due to the inhomogeneity of confining stresses or the reservoir's physical properties evidences that modelling of fracture propagation within the classical approaches where the pore pressure is not taken into account, can lead to significant errors in the estimation of the parameters of hydraulic fractures.

\section*{Acknowledgements}
The work is supported by RFBR (grant 16-01-00610) and President grant for support of Leading scientific schools (grant SSc-8146.2016.1).



\begin{thebibliography}{10}

\bibitem{Adachi}
J.~Adachi, E.~Siebrits, A.~Peirce, and J.~Desroches.
\newblock Computer simulation of hydraulic fractures.
\newblock {\em Int. J. Rock Mech. and Min. Sci.}, 44:739--757, 2007.

\bibitem{Biot1955}
M.~A. Biot.
\newblock Theory of elasticity and consolidation for a porous anisotropic
  solid.
\newblock {\em Journal of Applied Physics}, 26:182--185, 1955.

\bibitem{Biot1965}
M.~A. Biot.
\newblock Theory of propagation of elastic waves in a fluid-saturated porous
  solid. i.ii.
\newblock {\em Journal of the Acoustical Society of America}, 28:168--191,
  1956.

\bibitem{Carrier_Granet}
B.~Carrier and S.~Granet.
\newblock Numerical modeling of hydraulic fracture problem in permeable medium
  using cohesive zone model.
\newblock {\em Engng Fract Mech}, 79:312–--328, 2012.

\bibitem{Coussy}
O.~Coussy.
\newblock {\em Poromechanics}.
\newblock John Wiley \& Sons, 2004.

\bibitem{Detournay2016}
E.~Detournay.
\newblock Mechanics of hydraulic fractures.
\newblock {\em Ann. Rev. Fluid Mech}, 48:311--339, 2016.

\bibitem{GeertsmadeKlerk1969}
J.~Geertsma and F.~D. Klerk.
\newblock A rapid method of predicting width and extent of hydraulically
  induced fractures.
\newblock {\em J. Petrol. Tech.}, 21:1571--1581, 1969.

\bibitem{Golovin_Baykin_2016_Pore}
S.V. Golovin and A.N. Baykin.
\newblock Influence of pore pressure to the development of a hydraulic fracture
  in poroelastic medium.
\newblock {\em Int. J. Solids Struct.}
\newblock Under review.

\bibitem{Nordgren1972}
R.~P. Nordgren.
\newblock Propagation of a vertical hydraulic fracture.
\newblock {\em SPE J.}, 12:306--314, 1972.

\bibitem{PerkinsKern1961}
T.~K. Perkins and L.~R. Kern.
\newblock Widths of hydraulic fractures.
\newblock {\em J. Petrol. Tech.}, 13:937--949, 1961.

\bibitem{ShelBaikGol2014}
V.~V. Shelukhin, V.~A. Baikov, S.~V. Golovin, A.~Y. Davletbaev, and V.~N.
  Starovoitov.
\newblock Fractured water injection wells: Pressure transient analysis.
\newblock {\em International Journal of Solids and Structures}, 51:2116--2122,
  2014.

\bibitem{Z-K}
Y.~Zheltov and S.~Khristianovich.
\newblock On hydraulic fracturing of an oil-bearing reservoir.
\newblock {\em Izvestia AN SSSR}, ser. OTN, 1955.
\newblock (in Russian).

\end{thebibliography}

\end{document}